\documentclass[10pt,twocolumn,letterpaper]{article}
\usepackage{wacv}
\usepackage{times}
\usepackage{epsfig}
\usepackage{graphicx}
\usepackage{amsmath}
\usepackage{amssymb}
\usepackage{algorithm}
\usepackage[noend]{algorithmic}
\usepackage{booktabs,subcaption,amsfonts,dcolumn}
\usepackage{float}

\usepackage[pagebackref=true,breaklinks=true,letterpaper=true,colorlinks,bookmarks=false]{hyperref}

\wacvfinalcopy 

\def\wacvPaperID{319} 

\ifwacvfinal\fi
\begin{document}

\title{Progressive Multi-scale Consistent Network for Multi-class Fundus Lesion Segmentation}

\author{Along He\textsuperscript{1} , Kai Wang\textsuperscript{1},Tao Li\textsuperscript{1}, Wang Bo\textsuperscript{1}, Hong Kang\textsuperscript{1}, Huazhu Fu\textsuperscript{2}
\vspace{0.2cm}
\\
\textsuperscript{1}Nankai University\quad\textsuperscript{2}Institute of High Performance Computing (IHPC), Agency for Science, \\Technology and Research (A*STAR), Singapore}

\maketitle

\begin{abstract}

Effectively integrating multi-scale information is of considerable significance for the challenging multi-class segmentation of fundus lesions because different lesions vary significantly in scales and shapes. Several methods have been proposed to successfully handle the multi-scale object segmentation. However, two issues are not considered in previous studies. The first is the lack of interaction between adjacent feature levels, and this will lead to the deviation of high-level features from low-level features and the loss of detailed cues. The second is the conflict between the low-level and high-level features, this occurs because they learn different scales of features, thereby confusing the model and decreasing the accuracy of the final prediction. In this paper, we propose a progressive multi-scale consistent network (PMCNet) that integrates the proposed progressive feature fusion (PFF) block and dynamic attention block (DAB) to address the aforementioned issues. Specifically, PFF block progressively integrates multi-scale features from adjacent encoding layers, facilitating feature learning of each layer by aggregating fine-grained details and high-level semantics. As features at different scales should be consistent, DAB is designed to dynamically learn the attentive cues from the fused features at different scales, thus aiming to smooth the essential conflicts existing in multi-scale features. 
The two proposed PFF and DAB blocks can be integrated with the off-the-shelf backbone networks to address the two issues of multi-scale and feature inconsistency in the multi-class segmentation of fundus lesions, which will produce better feature representation in the feature space. Experimental results on three public datasets indicate that the proposed method is more effective than recent state-of-the-art methods.

\end{abstract}

\begin{figure}[!t]
\begin{center}
    \includegraphics[width=7.5cm]{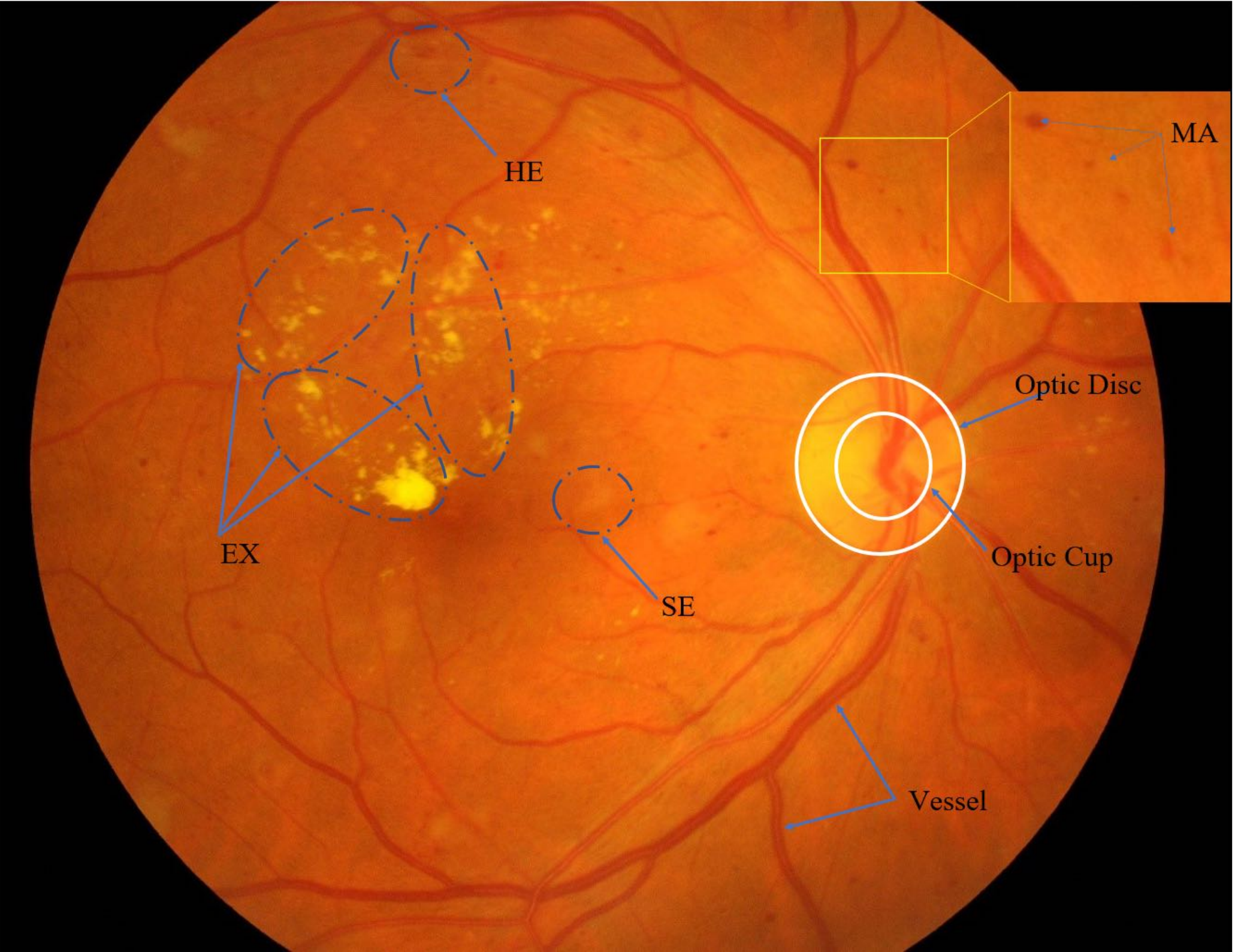}
\end{center}
\vspace{-0.5cm}
\caption{
   Example of fundus image from the IDRiD dataset. We marked some important biomarkers and four lesion types in the fundus images,  including optic disc, optic cup, blood vessel, HE (hemorrhages), MA (microaneurysms), SE (soft exudates), and EX (hard exudates). It can be seen from the fundus image that the lesions vary substantially in size, making the multi-class segmentation of fundus lesions difficult.}
\label{fig_1}
\vspace{-0.1cm}
\end{figure}

\section{Introduction}
\label{introduction}
Diabetic retinopathy (DR) is the most important manifestation of diabetic microangiopathy and one of the symptoms of diabetes mellitus. An estimated 451 million people were living with diabetes worldwide in 2017.
By 2045, the number is expected to increase to 693 million\cite{cho2018idf}.
Studies have shown that people with diabetes have an increased risk of developing DR and other cardiovascular and cerebrovascular diseases \cite{world2009global}.
The main pathological features of fundus lesions include microaneurysms (MA), hemorrhages (HE), soft exudates (SE) and hard exudates (EX).
In clinical practice, colored fundus images are widely used in the diagnosis of  DR because of their convenience and non-invasiveness to the human body.
These images provide some pathological information regarding the fundus, as shown in Fig. \ref{fig_1}.
Ophthalmologists usually diagnose fundus related diseases based on the presence of these lesions in colored fundus images.
However, owing to the limited number of ophthalmologists, screening a large number of fundus images is a significant burden.
The screening results are dependent on the ophthalmologist's experience, and this may lead to missed and misdiagnosed cases.
Therefore, automatic segmentation of these lesions from fundus images plays a vital role in the efficient screening and diagnosis of DR, significantly reduces the workload of ophthalmologists, and reduces the risk of missed diagnosis and misdiagnosis cases as well.

Recently, deep learning \cite{lecun2015deep} has made considerable progress in a wide range of tasks.
In particular, convolutional neural networks (CNNs) have been widely applied to various fields  of fundus image analysis \cite{li2021applications} and achieved satisfactory results, such as optic disc and optic cup segmentation\cite{fu2018joint}, vessel segmentation \cite{gu2019net}, and lesion segmentation \cite{guo2019seg}.
Although various methods of lesion segmentation have achieved good results, most of them focus on one or two types of lesion segmentation \cite{he2021incremental,dai2018clinical,chudzik2018microaneurysm}. 
Furthermore, the four types of fundus lesions cannot be segmented simultaneously, thereby limiting the diagnostic ability of the model for fundus screening.

As shown in Fig. \ref{fig_1}, the four types  of lesions  vary substantially in size, leading to the issue of inconsistent scales among lesions and making the multi-class segmentation of these lesions difficult.
The multi-scale feature fusion method is helpful to this issue.
As the multi-scale issue is ubiquitous in image segmentation tasks, several methods have been proposed to address this problem in natural image segmentation tasks, such as PSPNet \cite{zhao2017pyramid} and DeepLabv3+ \cite{chen2018encoder}.
However, these multi-scale feature fusion methods rely only on the high-level features to generate multi-scale features.
Although this enhances the ability to extract global context information, it ignores the details of the low-level features, which are crucial for small lesion segmentation.
At high levels of the encoder, the features contain more semantic cues but less spatial detailed cues, thus limiting the effects of multi-scale feature fusion, and are not suitable for the segmentation of small lesions.

As the popular models used in medical image segmentation tasks, U-Net \cite{chen2018encoder} and its variants \cite{ronneberger2015u,zhou2018unet++,li2018h} can utilize features from encoder by skip connections.
However, it only fuses the features from the same feature level, and the features are directly concatenated without distinction, resulting in two drawbacks. One is that they ignore the feature interaction of the adjacent layers and cannot fuse the  multi-scale features in the encoder.
The other is that the feature inconsistency of the fused features is ignored, increasing the difficulty of the final prediction.
These two overlooked aspects have severe negative impacts on the performance for the multi-class and multi-scale fundus lesion segmentation.

The key to solving this issue is the optimization of the fusion method for features extracted from multi-scale encoder layers.
Small lesions mainly rely on high-resolution features that have sufficient details. However, these low-level high-resolution features inevitably introduce background noises and need to fuse high-level context-aware information to guide the learning of low-level features. Unfortunately, features from different resolutions are conflict in scale and suppress low-level lesion features, which needs to further dynamically select useful features  from different resolutions. In this way, both detailed information and contextual information are maintained, and the background noises in the high-resolution feature maps can be effectively reduced.
This requires interactive learning and fusion of adjacent multi-scale features that contain high-level semantic information and low-level detailed information.

Based on this analysis, we introduce a progressive feature fusion (PFF) block to handle the challenging multi-scale issue, which is a  simple yet effective method for multi-scale fundus lesion segmentation.
We fully utilize the features from different encoding levels progressively, which has two advantages.
The first is realizing the fusion of multi-scale features without introducing other multi-scale feature extraction branches.
As the receptive fields of adjacent layers are different, lesions of different sizes can be handled simultaneously.
The second is reducing the influence of noises (background, optic cup, optic disc, and blood vessels) by utilizing the progressive learning of low-level features and high-level features, useless information can be filtered out as much as possible.

The features from various scales are different intrinsically, and to improve the consistency of the fused multi-scale features, we further propose a dynamic attention block (DAB), which is used to perform dynamic selection of the fused multi-scale features and adaptively learn the weights for different attentive mechanisms including channel-wise, spatial-wise, and point-wise mechanisms.
This makes multi-scale feature fusion more flexible and improves the consistency of multi-scale features.
By integrating PFF and DAB progressively with the off-the-shelf backbones to segment multi-scale fundus lesions in an end-to-end network, and it produce five feature maps, corresponding to four types of lesion maps and one background map.
The main contributions of our study are summarized as follows:

1.	We propose a simple yet effective PFF block to fuse the features from adjacent high-level and low-level layers, which enables progressive learning of high-level semantic information and low-level detailed information in each layer.
Therefore, it can strengthen the ability to preserve both detailed and contextual information simultaneously in encoder layers, and lesions of different sizes can be handled well.

2.	Further, we propose a DAB to realize dynamic feature selection via adaptive learning of different attentive mechanisms to solve the issue of conflicts in multi-scale features derived from different scales and improve the consistency of the fused features.

3.	Extensive experiments are conducted on three public datasets to verify the effects of the proposed PFF and DAB blocks. Compared with other state-of-the-art methods, our model achieves new state-of-the-art results in the task of fundus lesion segmentation.\footnote{The code will be available at \url{https://github.com/NKUhealong/PMCNet}}

This paper is organized as follows. Section \ref{related} provides a review in fundus lesion segmentation.
Section \ref{method} presents our proposed method in detail.
In Section \ref{experiments}, we evaluate the PMCNet comprehensively and compare it with other methods.
Conclusions and future work are given in Section \ref{Conclusion}.

\section{Related Work}
\label{related}

In this section, we briefly review the recent works on DR lesion segmentation, including single-class and multi-class lesion segmentation.

Lesion segmentation in fundus images is a challenging task because these lesions vary greatly in size, shape, location, color, and texture.
Owing to the development of deep learning algorithms, various methods based on CNNs have been used for lesion segmentation \cite{van2016fast,tan2017automated}.
For MA segmentation, a novel patch-based fully CNN was proposed in \cite{chudzik2018microaneurysm} to detect MA in fundus images, it can transfer the knowledge between different datasets.
To bridge the gap between image and diagnostic information, Dai et al. \cite{dai2018clinical} proposed a clinical report--guided CNN that leverages a small number of labeled fundus images in clinical reports to detect potential MA lesions.
For the segmentation of HE, Grinsven et al. \cite{van2016fast} proposed a  CNN-based method to improve the performance and accelerate the training by dynamically choosing misclassified negative images during the training stage and achieved good results.
For EX segmentation, Feng et al. \cite{feng2017deep} presented a method by adding short and long skip connections in the FCN architecture with class-balancing loss and achieved strong performance.
Guo et al. \cite{guo2020bin} proposed a bin loss function to deal with class imbalance in hard exudate segmentation, which can focus on hard-to-classify pixels and achieve good results.
High-resolution fundus images cannot be directly used as input because of the limited GPU memory, although it can be solved by down-sampling, while the detailed information is lost.
Alternatively, patches can be used to preserve details, but global context information is discarded because of the local receptive field, especially for lesions at the edge of patches.
Therefore, Yan et al. \cite{yan2019learning} proposed a network that integrates local and global features together, not only being aware of local details but also fully utilizing the global context information of fundus images, and this outperforms existing patch-based and full-image-based methods.

\begin{figure*}[!t]
	\centerline{\includegraphics[width=1\linewidth]{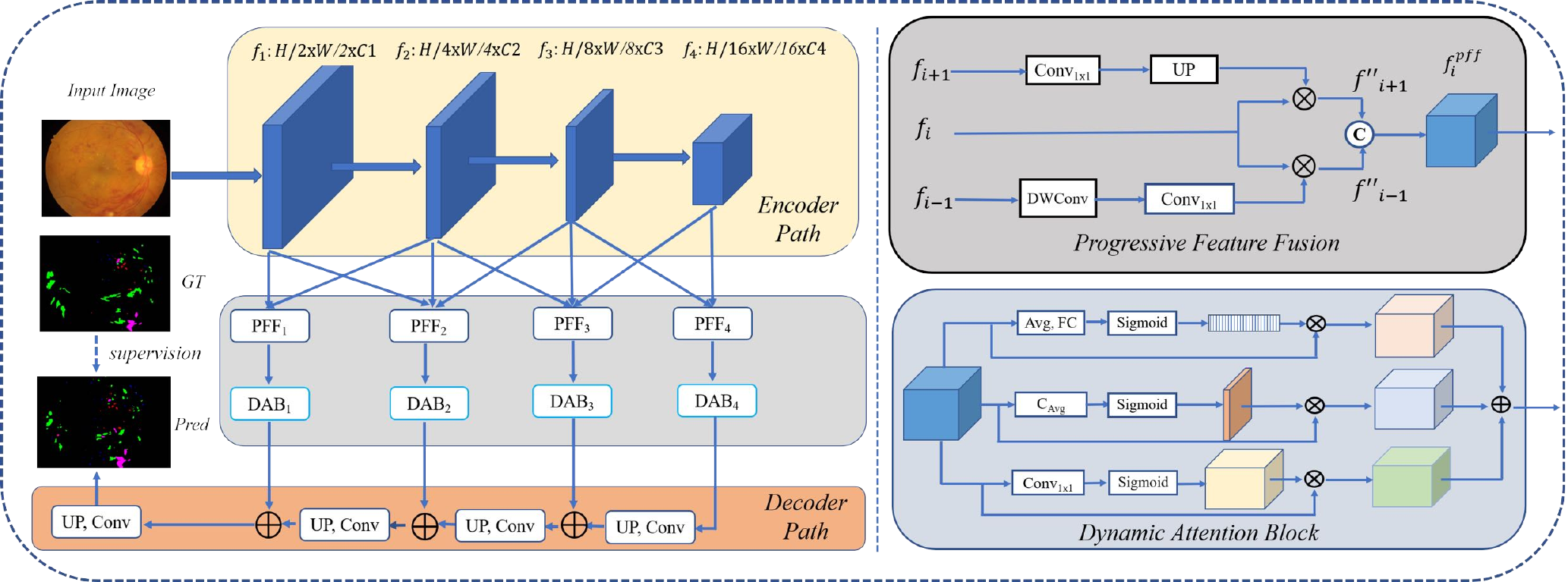}}
	\caption{Overall structure of the proposed PMCNet. It consists of four components: an encoder based on pre-trained models, a progressive feature fusion (PFF) block, a dynamic attention block (DAB), and a decoder path. ``GT'' and ``Pred'' denote the ground truth and prediction of the input fundus images, respectively. we use C1, C2, C3 and C4 to denote the number of feature channels in the four stages.}
	\label{fig_2}
\end{figure*}

Because pixel-level annotations for fundus images are very expensive, weakly supervised learning for DR lesion segmentation \cite{gondal2017weakly,gonzalez2020iterative} is a promising method.
The coarse bounding box label is a relatively good alternative solution, and in \cite{huang2020automated}, authors proposed a method to detect HE from coarsely labeled fundus images, which can significantly reduce the dependence on pixel-level label information.
To compensate for the lack of data, Playout et al. \cite{playout2019novel} proposed a multi-task network architecture that segments red and bright lesions simultaneously by using fundus images with image-level labels.
For multi-class fundus lesion segmentation, a multi-scale feature fusion method based on deep supervision was proposed in \cite{guo2019seg}, which can segment the four types of lesions simultaneously.

Although algorithms in lesion segmentation have made significant progress, there are still the following issues to be addressed.
First, most of the previous works focused on single-class lesion segmentation, and only a few studies focused on multi-class lesion segmentation.
The second is that existing works only fuse features from the same encoding layer to produce multi-scale features, which ignores the multi-scale role of features in the adjacent layers.
The third is that the multi-scale features are directly concatenated without distinction, which increases the difficulty of the final prediction owing to the conflict existing in different scales of features.

\section{Methodology}
\label{method}

In this section, we first provide a brief overview of the proposed network and then detail each component of the proposed method in the following subsections.

\subsection{Overview of PMCNet}

As shown in Fig. \ref{fig_2}, the proposed network follows the way of encoder-decoder, which consists of four components, including the encoder based on commonly  used CNNs pre-trained on ImageNet\cite{russakovsky2015imagenet}, the progressive feature fusion (PFF) block, the dynamic attention block (DAB), and the decoder.
For the encoder, we removed the fifth down-sampling stage, and the remaining layers were grouped into four stages by considering the feature maps with the same resolution as a stage.
The fundus image is fed into the encoder to extract low-level and high-level features gradually, and the input image is down-sampled four times. Thus, we denote the encoder part as  $E(x;\theta) = \{E_1,E_2,E_3,E_4\}$,  where $E_i (i=1,...,4)$ denotes the convolution, batch normalization, and activation layers in the $i$-th stage,   $\theta$ and $x$ denote the network parameters and input images, respectively.
The extracted side output feature maps from the last convolution layer in the $i$-th stage can be denoted as $ \{f_1,f_2,f_3,f_4\}$, which forms the vanilla multi-scale feature series. After the feature is extracted from the encoder path, the proposed PFF blocks progressively aggregate the adjacent high-level and low-level multi-scale features to produce enhanced feature representation in each layer and handle multi-scale lesions well.

To make multi-scale feature fusion more flexible and alleviate feature inconsistency among the fused multi-scale feature series, the proposed DAB will dynamically select useful features from different scales and refine the fused features to produce consistent multi-scale feature series with a dynamic learning strategy, which can adaptively learn the channel-wise, spatial-wise, and point-wise weights for each feature map in a triplet attention manner. In the decoder path, we up-sample the feature maps with a linear interpolation operation and reduce the number of feature channels gradually, and finally obtain the lesion segmentation probability maps with a 1 $\times$ 1 convolutional and softmax layer.

\subsection{Progressive Feature Fusion Block}
\textbf{\textit{Motivation}}: As is known to all, it is extremely important to maintain high-resolution features for segmenting small objects \cite{liu2020hrdnet} such as small lesions in fundus images.
This is because convolution and down-sampling operations will cause the loss of detailed information that is key to handling small lesions.
However, using high-resolution features directly may introduce too much background noises (i.e., the non-lesion pixels), because these low-level features are generated with a small receptive field and lack of global contextual information.
An intuitive solution is to increase the receptive field by dilated convolution \cite{chen2017deeplab} on high-resolution feature maps to enhance global semantic information.
Although the receptive field can be enlarged by different dilation rates, it also brings about an increase in computation cost and memory footprint because it adopts high-resolution feature maps.
Moreover, the dilated convolution also brings about the gridding effect \cite{liu2018receptive}, which is harmful to lesion segmentation.

Another solution is to use a UNet \cite{ronneberger2015u} structure that can perform feature fusion well through skip connections.
However, the skip connections in UNet only fuse features of the same spatial resolution on a single scale.
This limits the multi-scale feature extraction capability due to the lack of multi-scale information interaction and constraints among adjacent layers, leading to the deviation of high-level features from low-level features.
Therefore, the PFF block is proposed to effectively utilize the multi-scale information between adjacent feature scales gradually, with the aim of capturing fundus lesion features in different shapes and scales from low-level to high-level step by step, and thus, we term it as ``progressive feature fusion''.
The proposed PFF block can also be regarded as a feature constraint between adjacent layers, which means the features at a high level will not deviate from the features at a low level to maintain the overall consistency of the extracted features.
Unlike the commonly used multi-scale feature fusion that integrates all the multi-scale features at one time, our PFF block extracts multi-scale features from adjacent levels step by step, rather than only the same-level features of the encoder and the decoder being fused like UNet or its variants. The PFF block uses all levels of features in the feature extractor, rather than only the last level of features being used, such as PSPNet \cite{zhao2017pyramid} and DeepLabv3+ \cite{chen2018encoder} .

\textbf{\textit{Structure details}}: 
The structure of the PFF block shown in Fig. \ref{fig_2} is used four times to progressively integrate adjacent feature series, and each one has its own parameters.
Note that for the first and last PFF blocks, they only integrate two scales of features, as they are from the first and last side output layers. Three adjacent feature series were used for the rest of the PFFs.

Formally, for the $PFF_i$ block, there are two or three input feature scales:  high-level features $f_{i+1} (i=1,2,3)$, middle-level features$f_i$, and low-level features $f_{i-1} (i=2,3,4)$. Here, high-level and low-level features  are relative, for example, $f_2$ is high-level feature relative to $f_1$, but low-level feature relative to $f_3$.
For high-level features $f_{i+1}$, we first apply a 1 $\times$ 1 convolutional layer to change the channel number to the same number of $f_i$, and then, we upsample the feature maps to the same resolution as $f_i$.
We can get $f'_{i+1}$ by the following formula:
\begin{equation}
	f'_{i+1} = UP(Conv_{1\times1}(f_{i+1})), (i=1,2,3, 4),
\end{equation}
where $f'_{i+1}$ is the transformed high-level features. Because it comes from the high-level layers, it is rich in semantic information, which is able to guide the feature learning of the middle level, so that the fused features will retain the global information of the high level as much as possible to distinguish the easily confused  lesions.
$Conv_{1\times1}$ and $UP$ denote the 1 $\times$ 1 convolutional layer and 2 $\times$ upsample layer, respectively.

For low-level features $f_{i-1}$, we first perform a depth-wise convolutional layer \cite{howard2017mobilenets} with a stride of 2 to perform down-sampling, and the information of the four neighborhoods in the feature maps is retained in a weighted sum manner, which can avoid the loss of information by conventional max-pooling operation.

Then, a 1 $\times$ 1 convolutional layer is used to perform channel transformation, which changes the channel number to the same number of $f_i$. We can obtain $f'_{i-1}$ using the following formula:
\begin{equation}
	f'_{i-1} = Conv_{1\times1}(DWConv(f_{i-1})), (i=1,2,3,4),
\end{equation}
where $f'_{i-1}$ is the transformed low-level features, and it contains more detailed  lesion information that is beneficial for handling small lesions, which can be used to guide fine-grained feature learning, thus preserving the features of small lesions as much as possible.
$DWConv$ denotes a 2 $\times$ down-sample with a depth-wise convolutional layer.

After transforming the input high-level information and low-level information, we conduct the element-wise multiplication of  $f'_{i+1}$ and $f'_{i-1}$ with $f_i$, respectively:
\begin{equation}
	\left \{
	\begin{array}{ll}   f''_{i+1} =  f'_{i+1} \otimes f_i , i=1,2,3,4\\ 
		f''_{i-1} =  f'_{i-1} \otimes f_i , i=1,2,3,4,\\  
	\end{array}
	\right.
\end{equation}
where  $f'_{i+1}$ and $f'_{i-1}$ denote the enhanced high-level and low-level features, respectively, and $\otimes$ is the element-wise multiplication. Note that $f'_{0}$ and $f'_{5}$ are 0 because $f_{0}$ and $f_{5}$ does not exist in the feature series and they are set to 0.
The multiplication operation can strengthen the multi-scale features of lesions while reducing background noises.
Low-level features contain more detailed information, and thus small lesion details ignored by the middle level can be activated by multiplying with $f'_{i-1}$. 
High-level features contain more global semantic information, and lesion regions with weak semantic information in the middle-level feature scale can be strengthened by multiplying with $f'_{i+1}$. A feature visualization of the PFF block is illustrated in Fig. \ref{fig_3}.
It can be seen from the visualization results, compared with the middle-level features $f_i$, the low-level features $f_{i-1}$ contain more background noises, and the high-level features $f_{i+1}$ contain less detailed information.
After the fusion of low-level and middle-level features, the background noises in the low-level features $f_{i-1}$ can be effectively reduced in the enhanced low-level features $f''_{i-1}$.
After the fusion of high-level and middle-level features, the features of lesions in $f_{i+1}$ can be enhanced effectively in $f''_{i+1}$ owing to the introduction of detailed information provided by the middle-level features.
Therefore, by fusing adjacent levels of features with the proposed PFF block, more powerful multi-scale features can be generated by concatenating the enhanced high-level and low-level features for the decoder:

\begin{figure}[!t]
	\centerline{\includegraphics[width=1\linewidth]{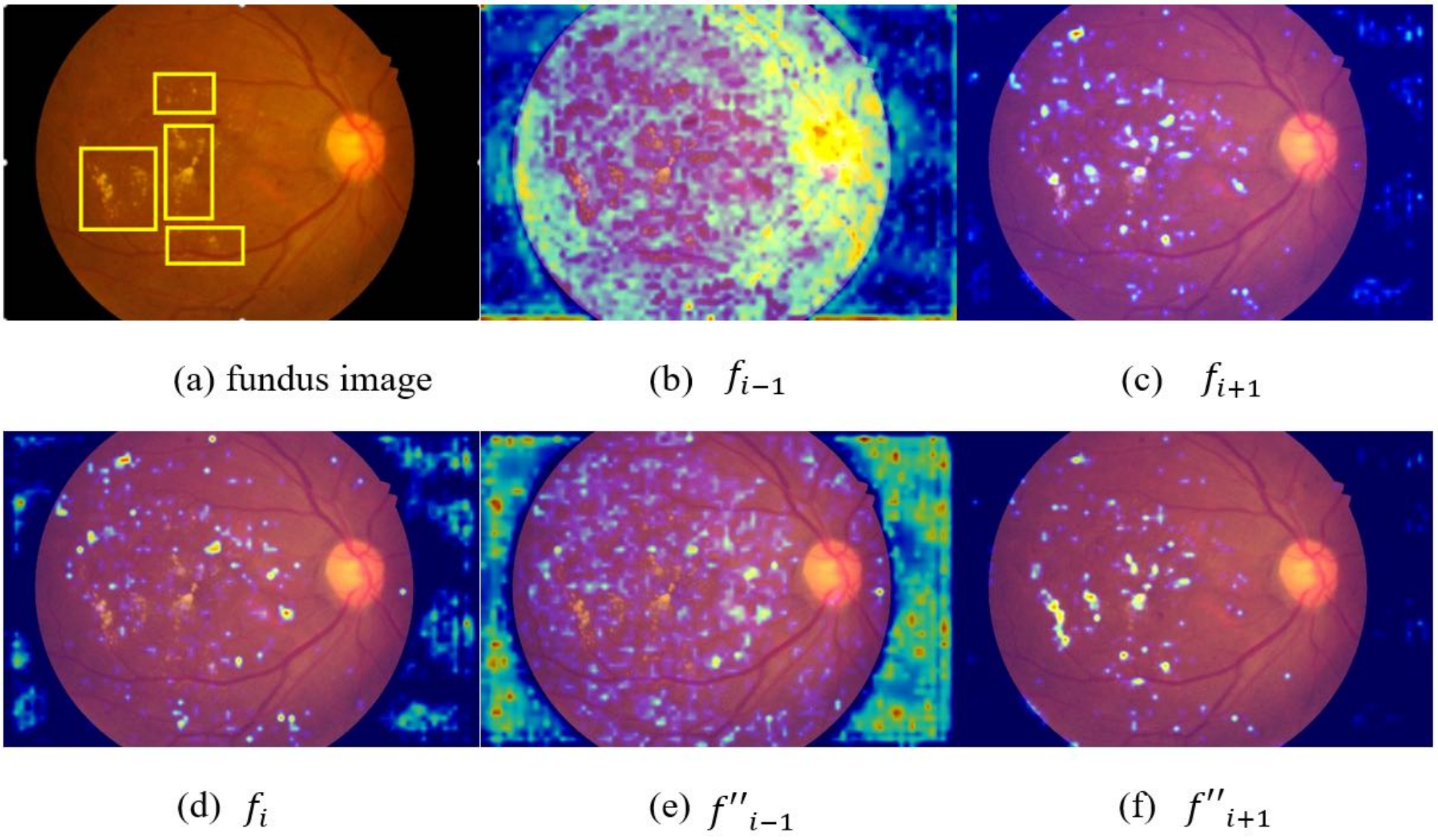}}
	\caption{ Heat map visualization around PFF, generated by Grad-CAM \cite{selvaraju2017grad}. From left to right in the first row are the fundus image, low-level features $f_{i-1}$, and high-level features $f_{i+1}$. The bottom row is the middle-level features $f_{i}$, enhanced low-level $f''_{i-1}$, and enhanced high-level features $f''_{i+1}$ ($i=3$ in this example). The yellow bounding boxes represent the lesion regions, and it can be seen that the model can capture the multi-scale lesion features well after PFF.  In (b) and (e), the lesion features around macular are not activated clearly, since they are low-level features and contain more background noises, thus the feature response is weakened, but not missing. This may be due to the fact that the response of background noise  is much higher than that of the lesion in low-level features, so the lesion response is overwhelmed, and they are well activated  in high-level, shown in (c) and (f).      }
	\label{fig_3}
\end{figure}

\begin{equation}
	f_{i}^{pff} = Concat(f''_{i-1}, f''_{i+1}), i=(1,2,3,4), 
\end{equation}
where $f_{i}^{pff}$ is the output of the $PFF_i$ block and $Concat$ is the concatenation operation.

Through the concatenation operation, detailed low-level information and high-level contextual information can be fused together to achieve multi-scale features progressively, which can enhance the feature representation ability for fundus lesions with different shapes and scales.

\subsection{Dynamic Attention Block}
\textbf{\textit{Motivation}:} As the features extracted from each level of the backbone are in different scales, simply aggregating these features will leads to feature inconsistency among different feature scales.
And treating all features equally results in information redundancy \cite{han2020ghostnet}.
Therefore, it is necessary to learn the importance of the fused multi-scale features dynamically to obtain an effective feature representation. Inspired by SENet \cite{hu2018squeeze}, 'channel squeeze \& spatial excitation' blocks \cite{roy2020squeeze} and multi-head attention \cite{vaswani2017attention}, we explicitly adopted the attention mechanism with the proposed dynamic attention block (DAB) to conduct effective feature selection via triplet attention. We aim to learn channel-wise, spatial-wise, and point-wise attention feature maps from three aspects to fully measure feature importance.
First, we perform channel attention by assigning different weights to different feature channels, and thus, different spatial positions within a feature channel share the same weight. This is suitable for selecting specific feature patterns in feature channels for the lesion segmentation task.
Second, we adopt spatial attention to deal with spatial position information.
The spatial location information is important for lesion segmentation.
Each feature map shares the same spatial attention weights, which highlights the spatial position of the lesion in a feature tensor.
Third, we further propose point-wise attention to perform point attention at each feature point, that is, different feature points have different point attention weights, which is useful for selecting the regions of small lesions.
Considering the different perspectives of each attention mechanism for lesion segmentation, we combine these three different attentions into triplet attention, and thus DAB is effective and suitable for performing dynamic feature selection and maintaining feature consistency.

\textbf{\textit{Structure details}}: The detailed DAB structure is shown in Fig. \ref{fig_2}, which takes the output of PFF as input.
First, channel-wise attention is calculated by the following formula:
\begin{equation}
	A^i_c = \sigma(FC(GAP(f_{i}^{pff}))\otimes f_{i}^{pff}, (i=1,2,3,4),
\end{equation}
where $A^i_c$ is the feature map generated by the channel-wise attention, $\sigma$ denotes the sigmoid function, $GAP$ is the global average pooling layer, $FC$ indicates the fully connected layer, $\otimes$ is the element-wise multiplication, and $f_{i}^{pff}$ is the input multi-scale features.
And then, spatial-wise attention is calculated by the following formula:
\begin{equation}
	A^i_s = \sigma(C_{Avg}(f_{i}^{pff}))\otimes f_{i}^{pff}, (i=1,2,3,4),
\end{equation}
where $A^i_s$ is the feature map generated by the spatial attention, and $C_{Avg}$ is the spatial-wise average pooling along the channel dimension to compress the features into a single channel.

\begin{figure}[!t]
	\centering
	\includegraphics[width=1.0\linewidth]{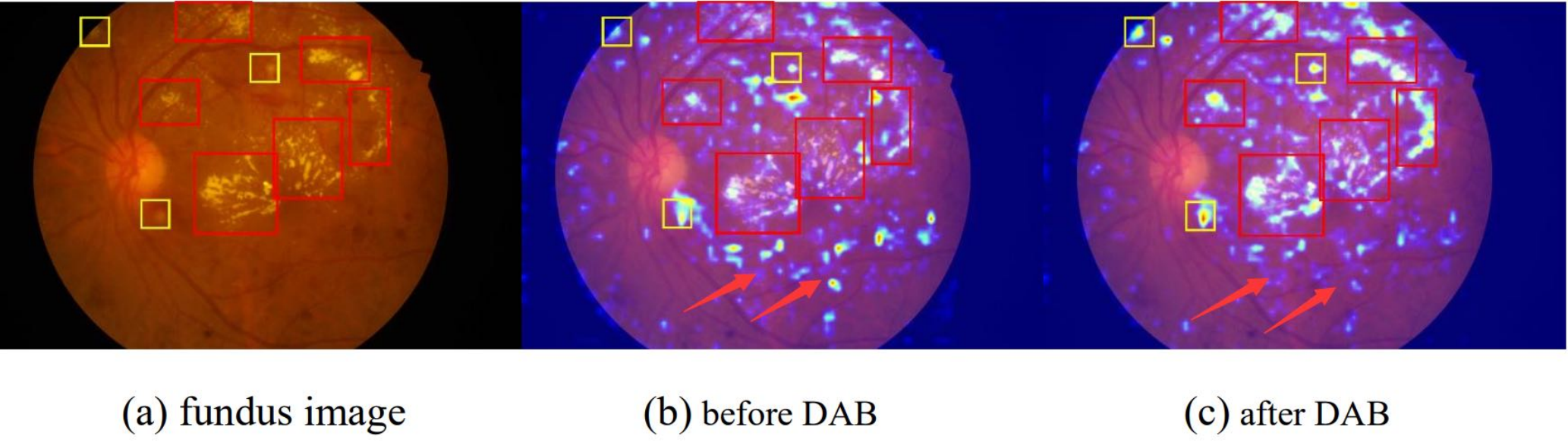}
	\caption{ Heat map visualization before and after DAB in the third stage. (a) Original fundus image, (b) The features before adopting DAB. (c) The features after adopting DAB. The yellow bounding boxes represent relatively small lesions of SE, the red bounding boxes represent relatively large lesions of EX, and the red arrows indicate the background, and it can be seen that the model can capture consistent multi-scale lesion features well and the most background noises can be reduced after adopting DAB. }   
	\label{fig_4}
\end{figure}

Finally,  we can obtain point-wise attention weights by using the following equation:
\begin{equation}
	W_i = \sigma(Conv_{1\times1}(f_{i}^{pff})), (i=1,2,3,4),
\end{equation}
where $W_i$ is the point-wise attention weights with the same shape as the input, which ranges from [0,1] and indicates the feature importance of each spatial point in each input feature map.
Therefore, it can be used for feature selection in a point-to-point manner and can improve the compatibility of multi-scale features from different feature scales.
Thus, point-wise attention feature maps can be obtained as follows:
\begin{equation}
	A^i_p = W_i \otimes f_{i}^{pff}, (i=1,2,3,4),
\end{equation}
where $A^i_p$ is the point-wise attention feature maps.
After obtaining the three attention feature maps, we can aggregate them by summation, and the formula can be described as follows:

\begin{equation}
	A^i = Conv_{1\times1}(A^i_c \oplus A^i_s  \oplus  A^i_p),
	\label{eq_4}
\end{equation}
where $A^{i}$ is the calibrated consistent attention features of the DAB, $Conv_{1\times1}$ is used to reduce the feature channels followed by the ReLU activation layer, and $\oplus$ is the element-wise addition.
After dynamic selection, redundant features are effectively reduced, and the gap between multi-scale features can be further reduced by weight reallocation.
The feature visualization of the DAB block is illustrated in Fig. \ref{fig_4}. We can see from the visualization results that the model focus on more discrete lesion dots and produces inconsistent features due to conflicts among multi-scale features before adopting DAB. After adopting DAB block, we can see that the resulting feature maps are more consistent, and can focus on  lesion information. Meanwhile, the background information is also effectively reduced as indicated by the red arrows.

As shown in Fig. \ref{fig_2}, the proposed PFF and DAB are integrated progressively. These blocks are repeatedly used in the four feature levels to gradually extract multi-scale features and make the model perceive multi-scale features, instead of integrating all feature scales simultaneously, so we call it progressive, and they can be integrated with the existing backbone networks.

\subsection{Decoder Path}
In the decoder path, we leverage the consistent multi-scale feature series $ \{A^1,A^2,A^3,A^4\}$ for decoding. We use linear interpolation for up-sampling and two consecutive 3$\times$3 convolutional layers followed by ReLU activation for feature learning.
Inspired by the design principle of UNet \cite{ronneberger2015u}, we add the features of the corresponding output from the encoder to the corresponding up-sampling stage:

\begin{equation}
	D_{i}=Conv_{3\times3}(UP(D_{i-1}) \oplus A^{4-i}), (i=1,2,3),
	\label{eq_7}
\end{equation}
where $D_i$ denotes the output features of each decoder stage, $D_0 = A^4$,   $\oplus$ denotes the element-wise addition, and $Conv_{3\times3}$ is the 3$\times$3 convolutional layer followed by  ReLU activation layer.

In the final stage, the feature maps are resized to the same resolution as the input, using a 1$\times$1 convolution layer followed by the softmax activation function to map them to five channels, corresponding to four types of lesion maps and one background map, respectively.

\subsection{Loss Function}
\label{loss}
In the training stage, we use cross-entropy loss as the objective function, which is defined as follows:

\begin{equation}
	L_{ce}(y,g)=-\sum_{i=1}^{k}{g_i \cdot log(y_i)}.
\end{equation}
where $g$, $y$, and $k$ are the ground truth, predicted lesion maps of the input fundus images, and the number of lesion types, respectively.

\section{Experiments}
\label{experiments}
In this section, we introduce the commonly used benchmark datasets for fundus lesion segmentation, implementation details, and evaluation metrics, explain the comprehensive experiments conducted to show the effectiveness of our proposed method, and provide qualitative and quantitative results. We also compare the results of the proposed method with those of the state-of-the-art methods.

\subsection{Benchmark Datasets}
\textbf{DDR Dataset} \cite{li2019diagnostic}: This dataset is a general high-quality dataset, which contains 757 color retinal fundus images with pixel-level annotation for multiple fundus lesion segmentation. There are 570 images with MA, 239 images with SE, 601 images with HE, and 486 images with EX.
For the dataset split, there are 383 images for training, 149 images for validation, and 225 images for testing.

\textbf{IDRiD Dataset} \cite{porwal2018indian}: This dataset consists of 81 images with the resolution of 2848$\times$4288, and pixel-level annotations of EX, HE, MA and SE are provided in this dataset. Each image may contain one or more fundus lesion labels. There is an official separation of the training and test sets for this dataset, 54 images
for training and the rest 27 images for testing.

\textbf{E-ophtha Dataset} \cite{decenciere2013teleophta}: The E-ophtha dataset is composed by two parts: e-ophtha EX and e-ophtha MA. Each part contains pixel-level lesion annotation, the e-ophtha-MA dataset contains 148 images with MA, and the e-ophtha-EX dataset contains 47 images with exudates. Considering the multi-class lesion segmentation task, we selected 21 images that contained both hard exudates and microaneurysm annotations from this dataset. In our experiments, we used 11 fundus images for training and the remaining 10 images for testing, as suggested in \cite{guo2019seg}. 

\subsection{Implementation Details} 
In the experiments, we applied horizontal flips, vertical flips, and random rescales as a form of data augmentation to reduce overfitting and improve the generalization ability of the model. The network was trained with the Adam optimizer \cite{kingma2014adam} to learn the weights. The number of feature channels for the four stages of MobileNet, VGG16, ResNet50 and EfficientnetB0 are [32,64,128,256], [32,128,256,512], [32,64,256,512] and [32, 96, 144, 240], respectively. The spatial resolution of the original training images is too large, and thus, we resize the images to reduce GPU memory consumption. The models are trained and tested separately on the three datasets. Specifically, for IDRiD and E-ophtha datasets, we resized each image and its ground truth to 960$\times$1440 (height$\times$width). Due to limited training images on the E-ophtha dataset, we augmented this dataset ten times using horizontal flips, vertical flips, and random rescales. For the DDR dataset, we resize them to 1024$\times$1024 (height$\times$width), as suggested in \cite{guo2019seg}.
We set the initial learning rate to 0.0001 and reduce the learning rate by the ``Poly'' strategy used in \cite{chen2017rethinking}, and the batch size is 1 due to limited GPU memory. The loss function is pixel-wise cross-entropy. Our framework was implemented with Keras using TensorFlow as the backend. All experiments were performed on an NVIDIA GTX 1080Ti GPU.

\subsection{Evaluation Metrics}
We used the area under the curve (AUC) of precision-recall to measure the performance of the proposed method, which is the same as the IDRiD challenge \cite{porwal2018indian}. And the mean AUC (mAUC) in \cite{guo2019seg} was employed to evaluate the overall performance of the proposed methods in a multi-class lesion segmentation task, and mAUC is the mean value of AUCs for different types of lesions.
In addition, to better measure the performance of the model, we further adopted IoU and Dice score as evaluation metrics.

\begin{table}[!btp]
\centering
\begin{tabular}{|l|c|c|c|c|}    
\hline
Method & Backbone  & 10\% lab& 20\% lab\\
\hline\hline
Supervised & 3D ResNet-18  & 66.75 & 75.79\\
\hline\hline
DMPCT~\cite{zhou2018semi}  &  2D ResNet-101 & 63.45& 66.75\\
DCT~\cite{qiao2018deep} (2v) & 3D ResNet-18 & 71.43& 77.54\\
TCSE~\cite{li2018semi} &  3D ResNet-18 & 73.87 & 76.46\\
\hline\hline
Ours (2 views) &  3D ResNet-18 &\textbf{75.63} & \textbf{79.77}\\
Ours (3 views) &  3D ResNet-18 &\textbf{77.55} & \textbf{80.14}\\
Ours (6 views) &  3D ResNet-18 &\textbf{77.87} & \textbf{80.35}\\
\hline
\end{tabular}
\vspace{-0.2cm}
\caption{
    Comparison to other semi-supervised approaches on NIH dataset (DSC, \%).  Note that we use the same backbone network as~\cite{li2018semi}~\cite{qiao2018deep}. Here, ``2v" means two views. For our approach, the average of all single views' DSC score is reported for a fair comparison, not a multi-view ensemble. ``10\% lab"  and ``20\% lab" mean the percentage of labeled data used for training.
}
\label{Tab:semiTE}
\vspace{-0.6cm}
\end{table}

\subsection{Ablation Studies}
To verify the effectiveness of our proposed PFF and DAB blocks, we conducted ablation studies to better understand the impact of each component based on the lightweight MobileNet \cite{howard2017mobilenets},  and we also adopt the skip connections like UNet in the four stages to fuse features of the same spatial resolution on a single scale. Finally, we demonstrate the generalization ability of the proposed module on different backbones.

\textbf{Analysis of PFF Block:}
We argue that low-level detailed information and high-level semantic information are both important for fundus lesion segmentation. Therefore, it is beneficial to use these two types of information simultaneously for the segmentation of small lesions with complex context. We investigated the effectiveness of the PFF block in Table \ref{table_1} on the IDRiD dataset and compared different variants of PFF: fusion only with adjacent low-level (high resolution, named PFF\_low) features, fusion only with adjacent high-level (low resolution, named PFF\_high) features,  fusion with adjacent low-level and high-level features by summation (PFF\_sum), the proposed PFF that aggregates both of them by multiplication in a progressive manner (PFF) and without progressive manner  (vanilla fusion, a common feature fusion in computer vision  \cite{zhao2017pyramid}, \cite{chen2018encoder}. In this manner, only the last two feature scales were concatenated without the progressive way, ignoring the high-resolution feature maps).

For PFF\_low, we aim to obtain more detailed information for the detection of small  lesions. As shown in Table \ref{table_1}, after we adopt  PFF\_low, the performance for EX, MA, and SE segmentation can be improved, especially for SE, improved by 3.67\% in AUC, which demonstrates that integrating low-level detailed features will enhance the small lesion detection ability of the model.
For HE, the segmentation performance is degraded because of the similarity between MA and HE, and it requires more multi-scale contextual information.

Soft exudates (SE) and hard exudates (EX) are similar in structure and texture; thus, global context information is needed to better distinguish them. As shown in Table \ref{table_1}, after adopting the PFF\_high, the segmentation ability of the model for SE is further improved, and the experimental results verified that the fusion of high-level semantic information can provide more global features of lesions and strengthen the context reasoning ability of the model for similar fundus lesions.

For down-sampling the feature map size, a conventional convolution with a stride of 2 could achieve it, but it will introduce a little more parameters and computation cost, and thus we used depth-wise convolution for down-sampling. Different from Max Pooling, we make a weighted sum of the four neighborhoods in a learnable way to ensure that the information within the four neighborhoods can be effectively fused. We also conducted ablation studies to verify the influence of down-sampling in Table \ref{table_1}, which show that depth-wise convolution performed best.

Compared with baseline+PFF\_sum, the results of the proposed PFF (DWConv) block indicate that the performances for EX, HE, and SE  will be further improved, which shows that the proposed PFF block can provide more useful multi-level information for different lesion scales than the summation method. These results also demonstrate that low-level and high-level features are both important.

For vanilla fusion, our results reveal that the segmentation performance drops significantly compared with PFF(DWConv), which demonstrates that a progressive manner will enhance the multi-scale feature representation ability and make full use of the multi-scale feature series. Compared with the baseline, the performance of MA decreases slightly even though the mAUC improved by 2.25\%. This is caused by the conflict and inconsistency of multi-scale features, and needs to be further investigated.

\begin{table*}
	\center
	\caption{Ablation studies of the proposed blocks on IDRiD dataset. The AUC and mean AUC (mAUC) are reported in this table. The baseline refers to MobileNet and the skip connections are used to fuse features of the same spatial resolution.}
	\setlength{\tabcolsep}{2pt}
	\renewcommand\arraystretch{1.2}
	\begin{tabular}{p{122pt}|p{41pt}p{41pt}p{41pt}p{41pt}|p{41pt}|p{40pt}}
		\hline
		Method&                EX&   HE&      MA&     SE&    mAUC&    $\Delta$ /\%\\
		\hline
		baseline                    &83.47&59.33&33.35&56.53&58.17& -\\
		
		baseline+PFF\_low             &83.85&58.96&33.69&60.20&59.18& $\uparrow$1.01\\
		baseline+PFF\_high            &83.45&56.35&34.79&61.73&59.08& $\uparrow$0.91\\
		baseline+PFF\_sum             &84.22&56.66&35.59&60.67&59.29& $\uparrow$1.12\\
		baseline+vanilla fusion       &83.41&59.77&33.90&56.79&58.47& $\uparrow$0.30\\
		baseline+PFF(MaxPool) &84.01&61.70&32.31&61.76&59.95&$\uparrow$1.78\\
		baseline+PFF(Conv)    &83.92&61.30&33.09&62.20&60.13&$\uparrow$2.01\\
		baseline+PFF(DWConv)            &84.30&\textbf{62.10}&33.11&62.22&60.42&$\uparrow$2.25\\
		
		baseline+DAB             &84.53&57.37&\textbf{38.89}&62.94&60.93&$\uparrow$2.76\\
		baseline+PFF+SE        &83.98 &60.86&35.61&63.29&60.94&$\uparrow$2.77\\
		baseline+PFF+CBAM       &83.63&60.36&33.84& 64.44&60.57&$\uparrow$2.40\\
		baseline+PFF+DAB(ours)       &\textbf{86.70}&61.16&38.51&\textbf{65.56}&\textbf{62.98}& $\uparrow$\textbf{4.81}\\
		\hline
	\end{tabular}
	\label{table_1}
\end{table*}
\begin{figure}[h]

	\subfloat[fundus]{\includegraphics[width=1.62in]{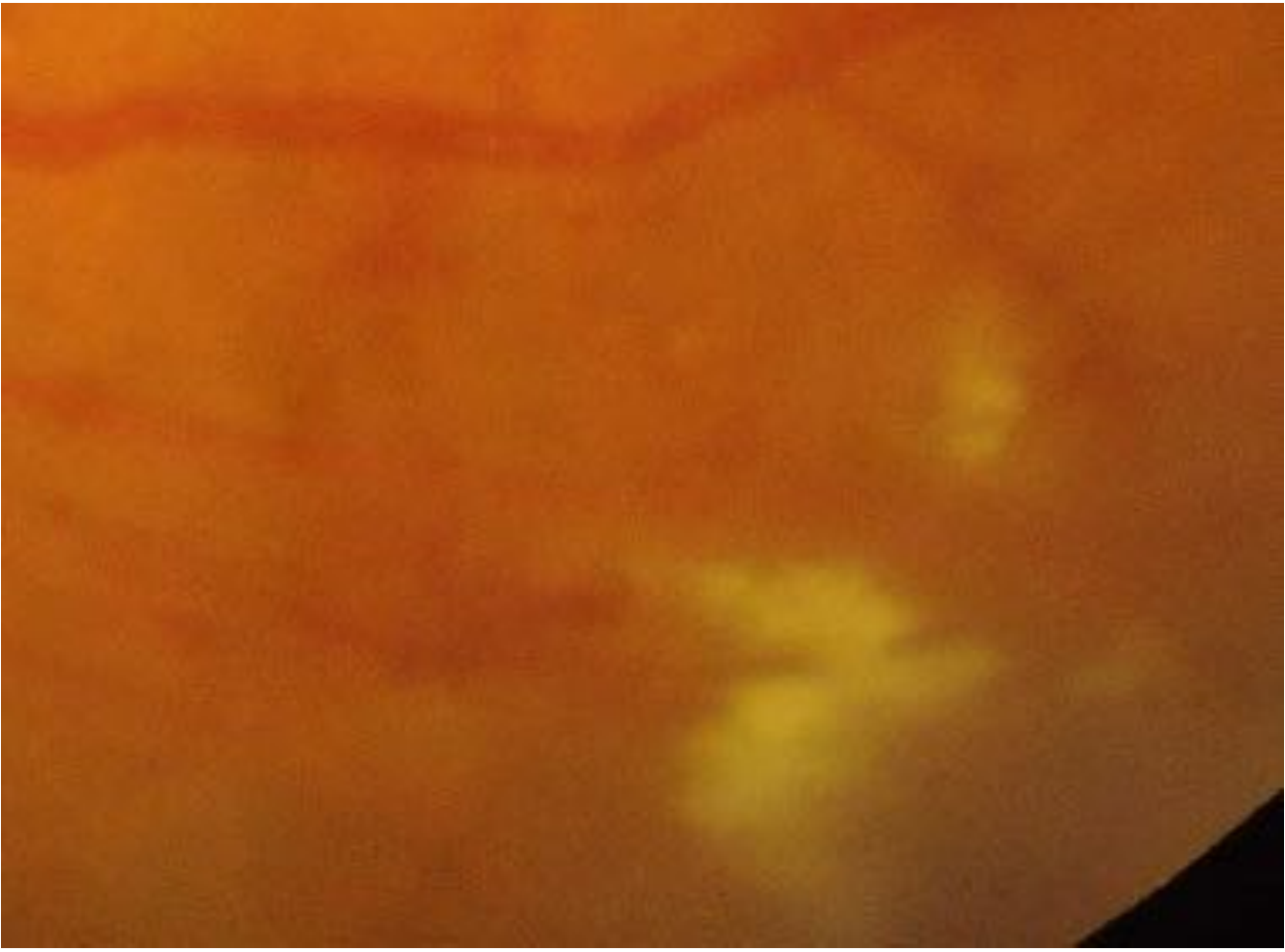}}
	\subfloat[GT]{\includegraphics[width=1.6in]{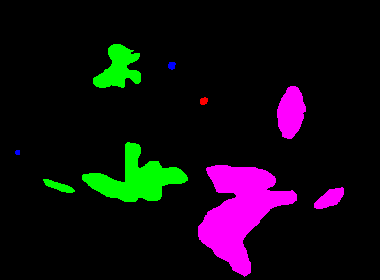}}
	
	\subfloat[ResNet50]{\includegraphics[width=1.6in]{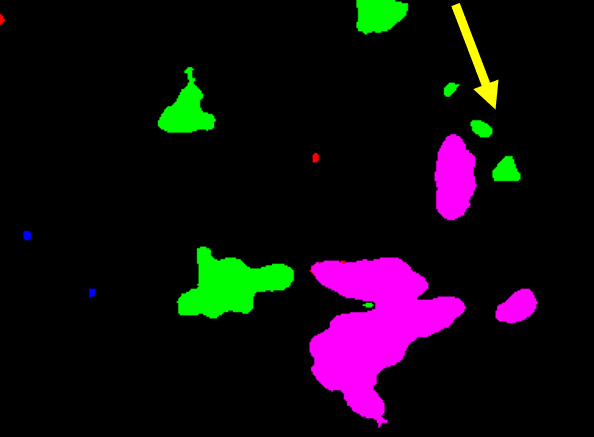}}
	\subfloat[ResNet50+PFF+DAB]{\includegraphics[width=1.6in]{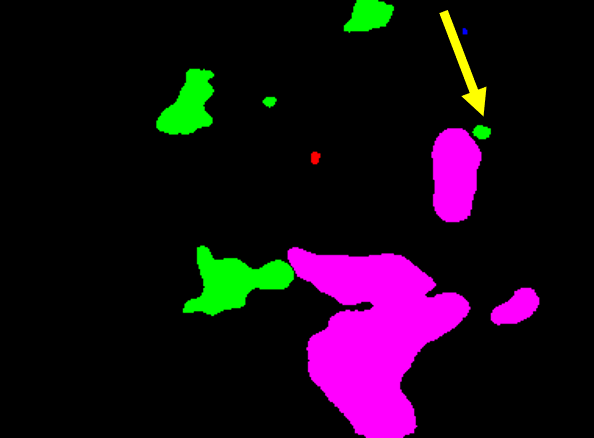}}
	
	\subfloat[VGG16]{\includegraphics[width=1.6in]{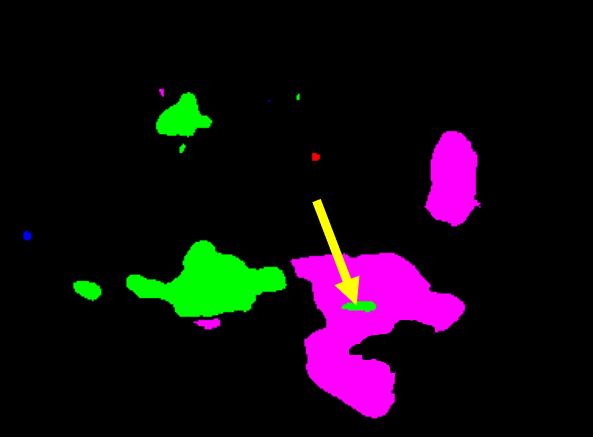}}
	\subfloat[VGG16+PFF+DAB]{\includegraphics[width=1.6in]{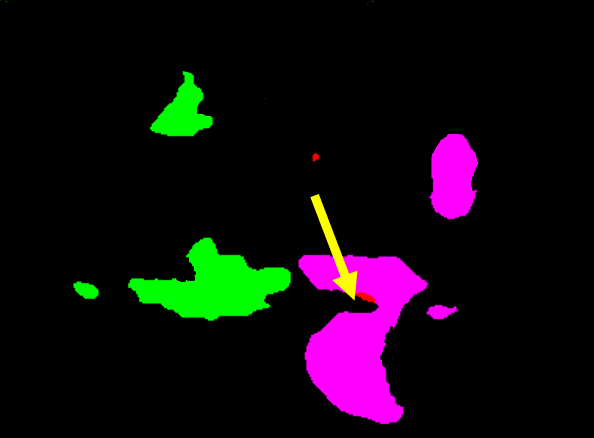}}
	
	\subfloat[EfficientnetB0]{\includegraphics[width=1.6in]{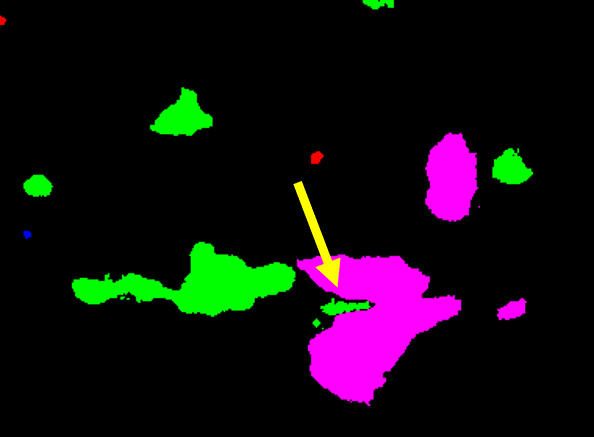}}
	\subfloat[EfficientnetB0+PFF+DAB]{\includegraphics[width=1.6in]{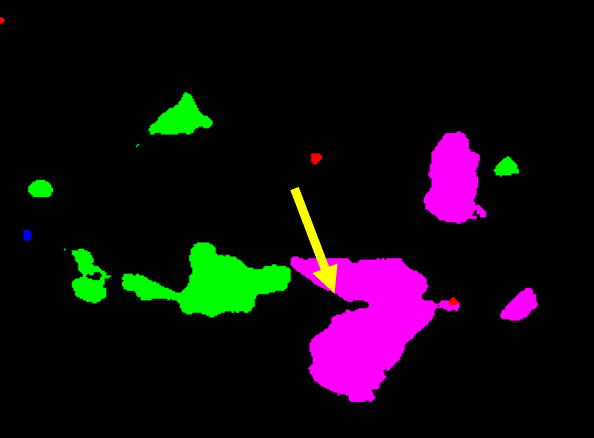}}
	
	\caption{Qualitative results of different backbones incorporated with the proposed PFF and DAB blocks. Red, green, blue, and pink markings denote EX, HE, MA, and SE, respectively. The yellow arrows indicate the differences between the baseline and proposed methods. }
	\label{fig_5}
\end{figure}
 
\begin{figure*}[h] 
	\centering
	\subfloat[fundus]{\includegraphics[width=1.69in]{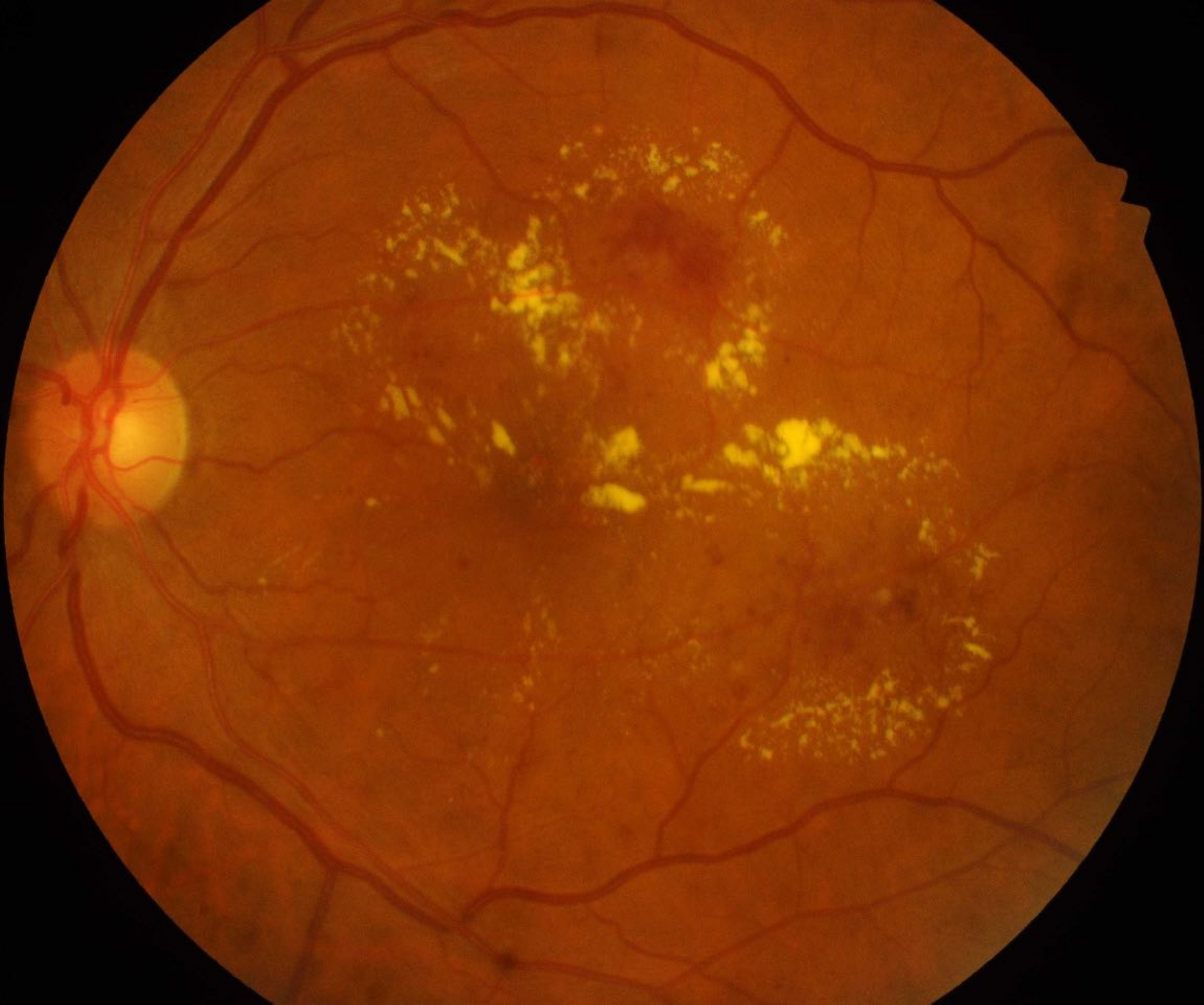}}
	\subfloat[GT]{\includegraphics[width=1.69in]{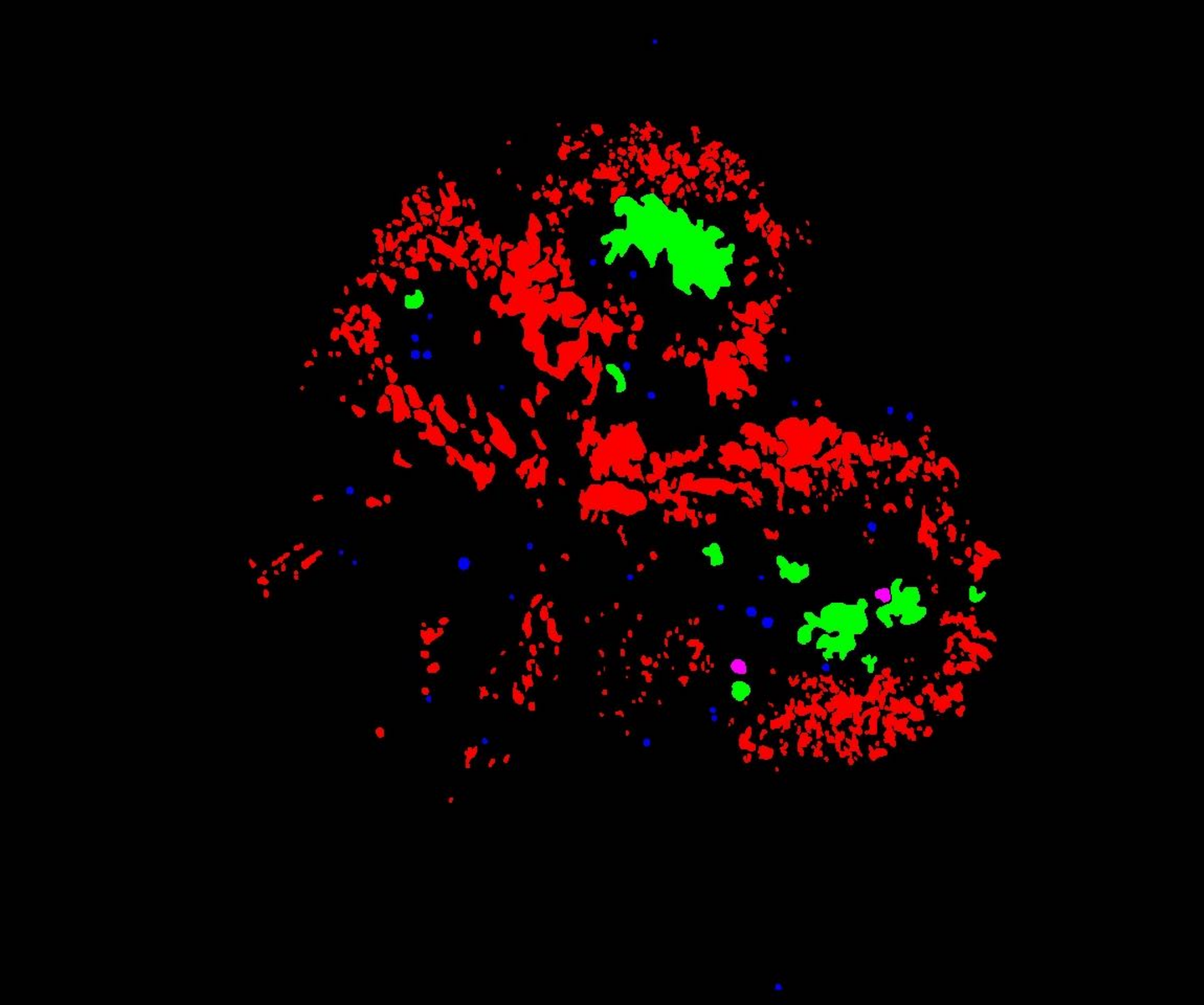}}
	\subfloat[UNet]{\includegraphics[width=1.69in]{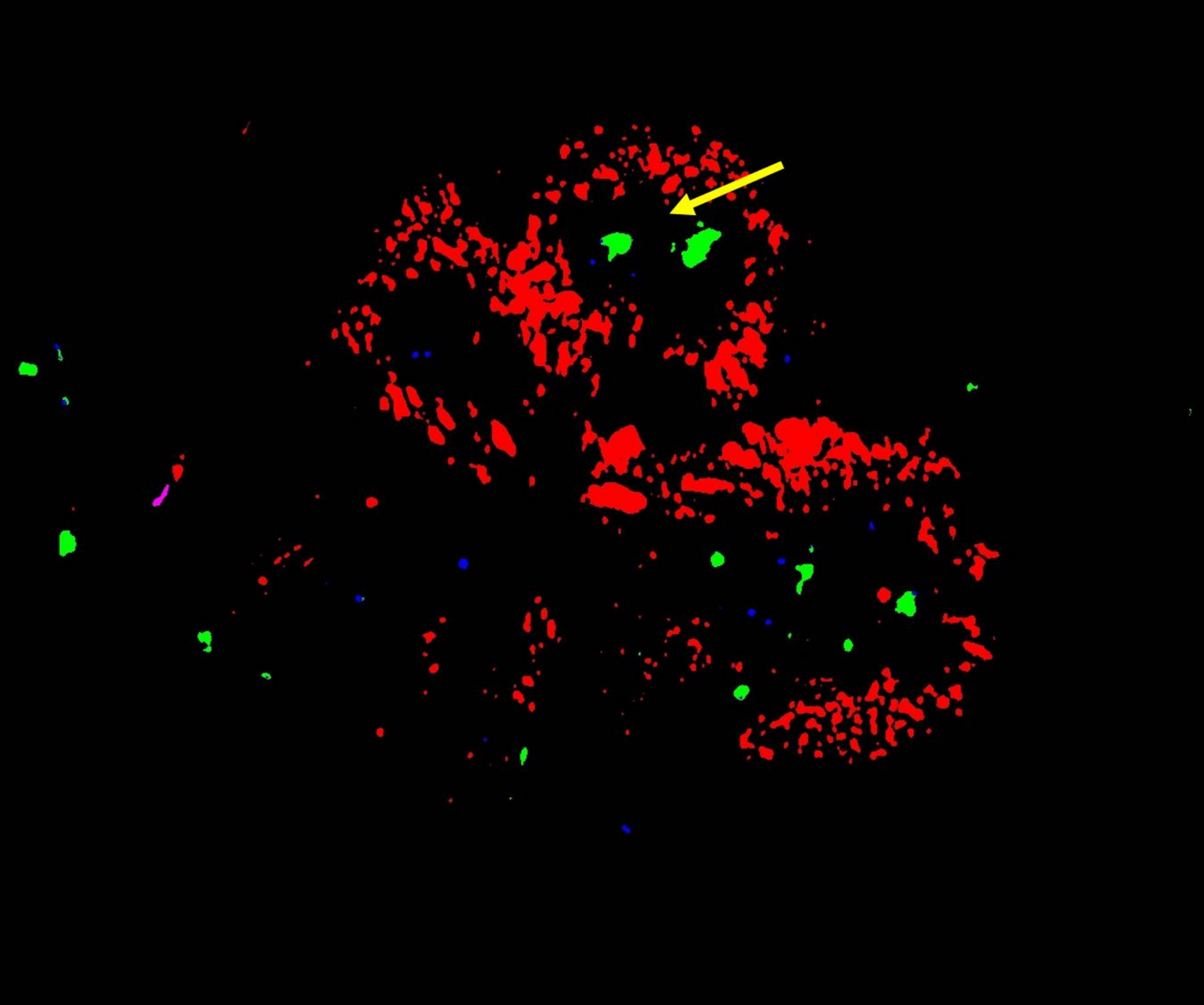}}
	\subfloat[ResUNet]{\includegraphics[width=1.69in]{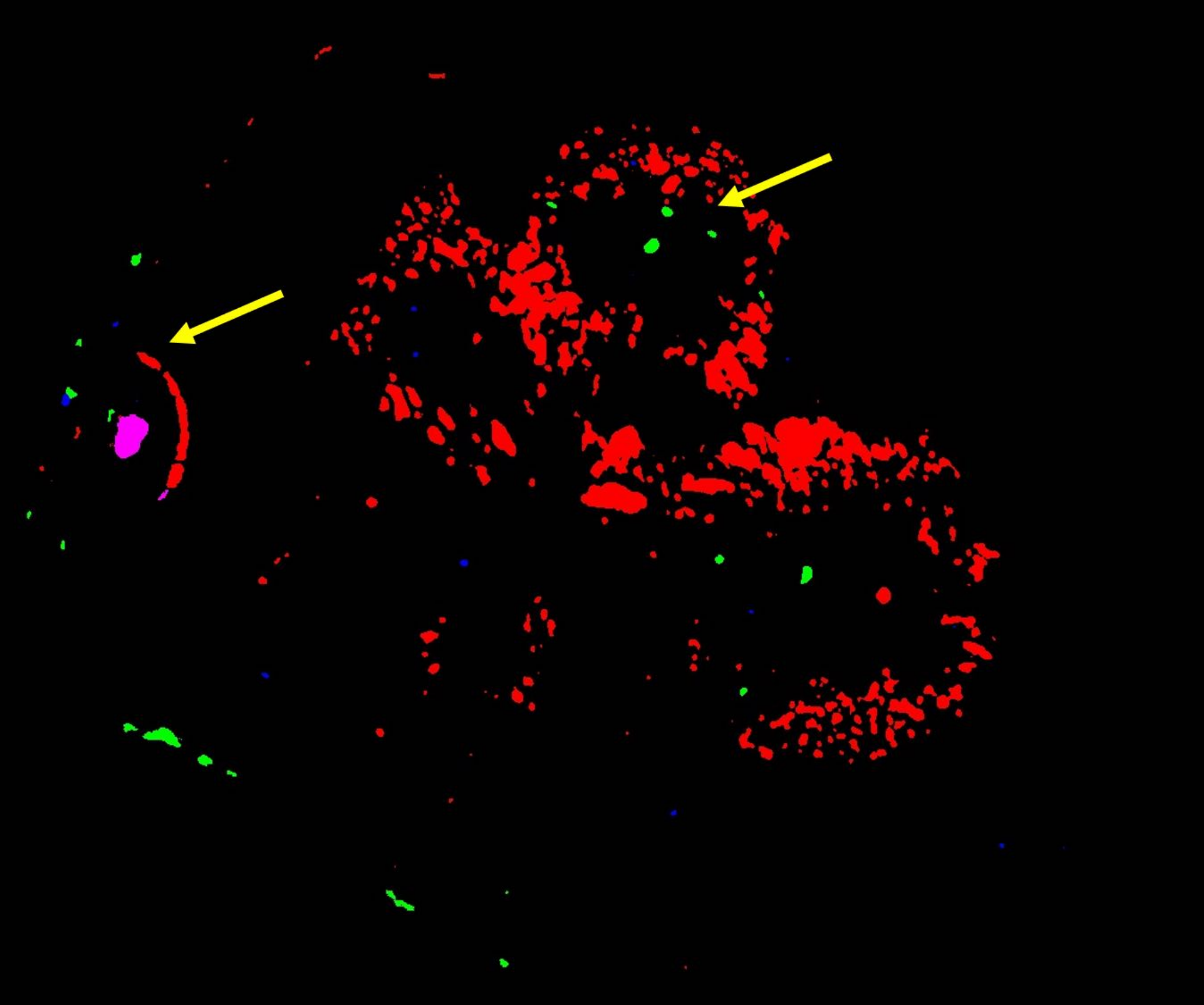}}
	
	\subfloat[DenseUNet]{\includegraphics[width=1.69in]{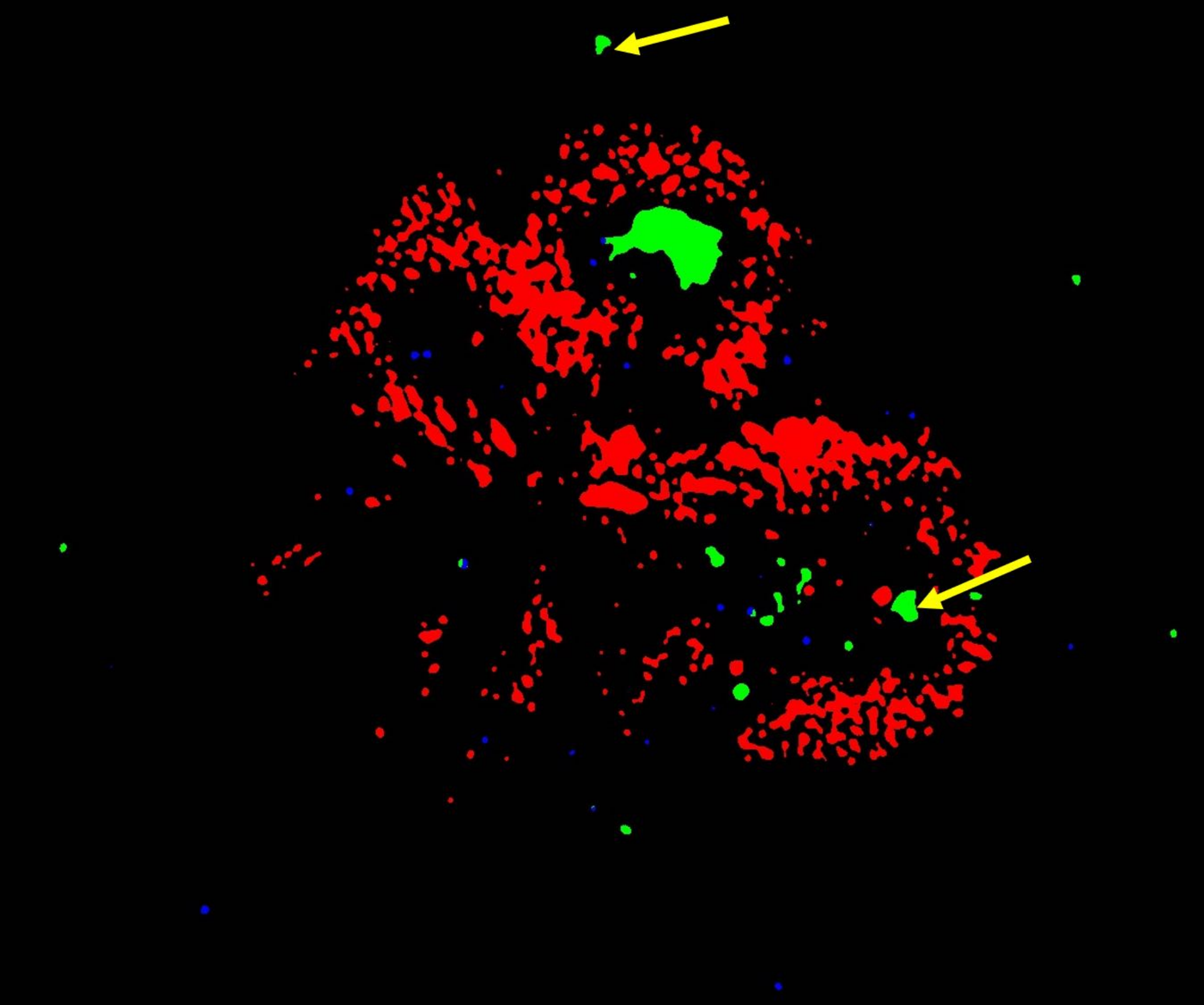}}
	\subfloat[UNet++]{\includegraphics[width=1.69in]{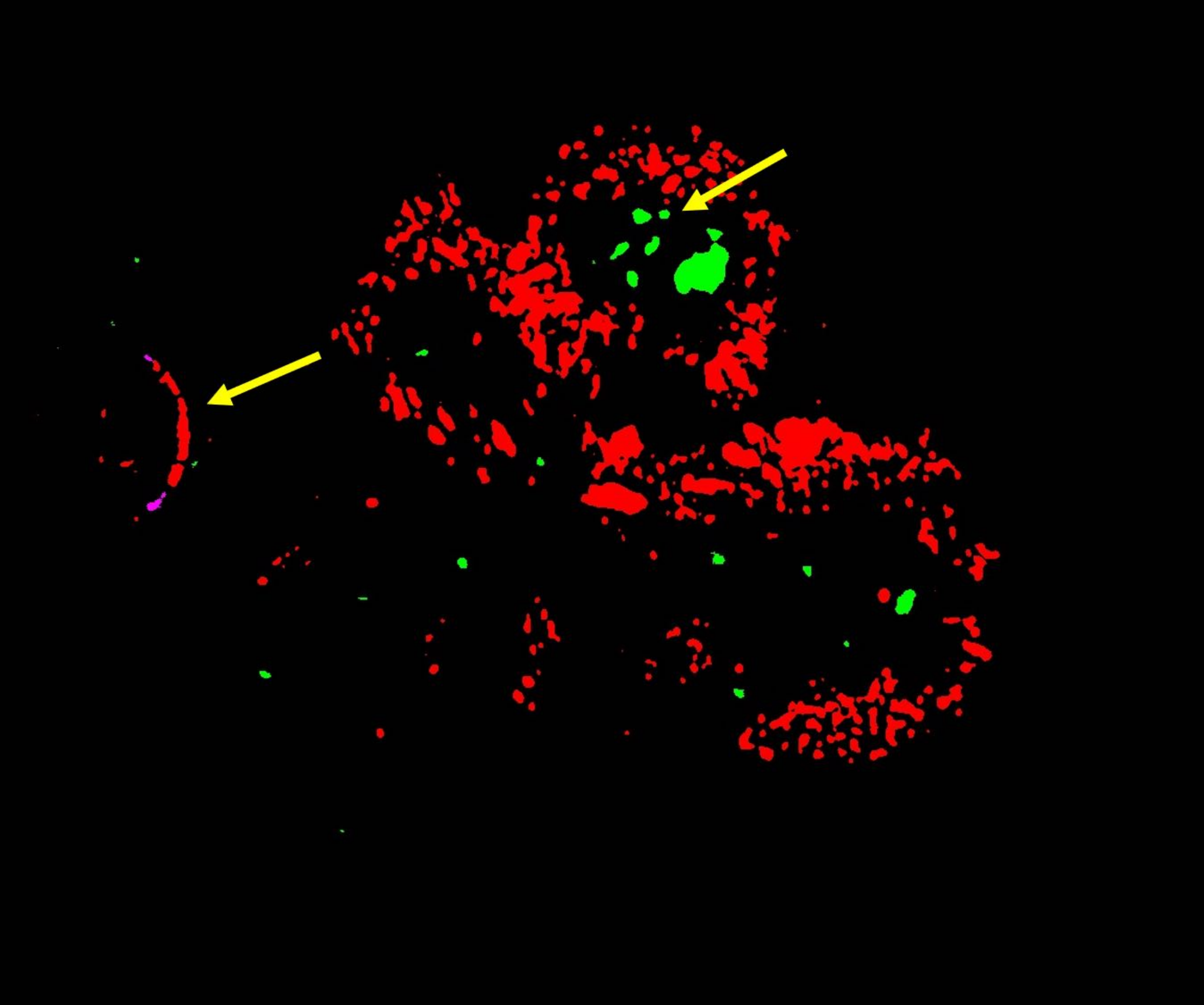}}
	\subfloat[DeepLabv3+]{\includegraphics[width=1.69in]{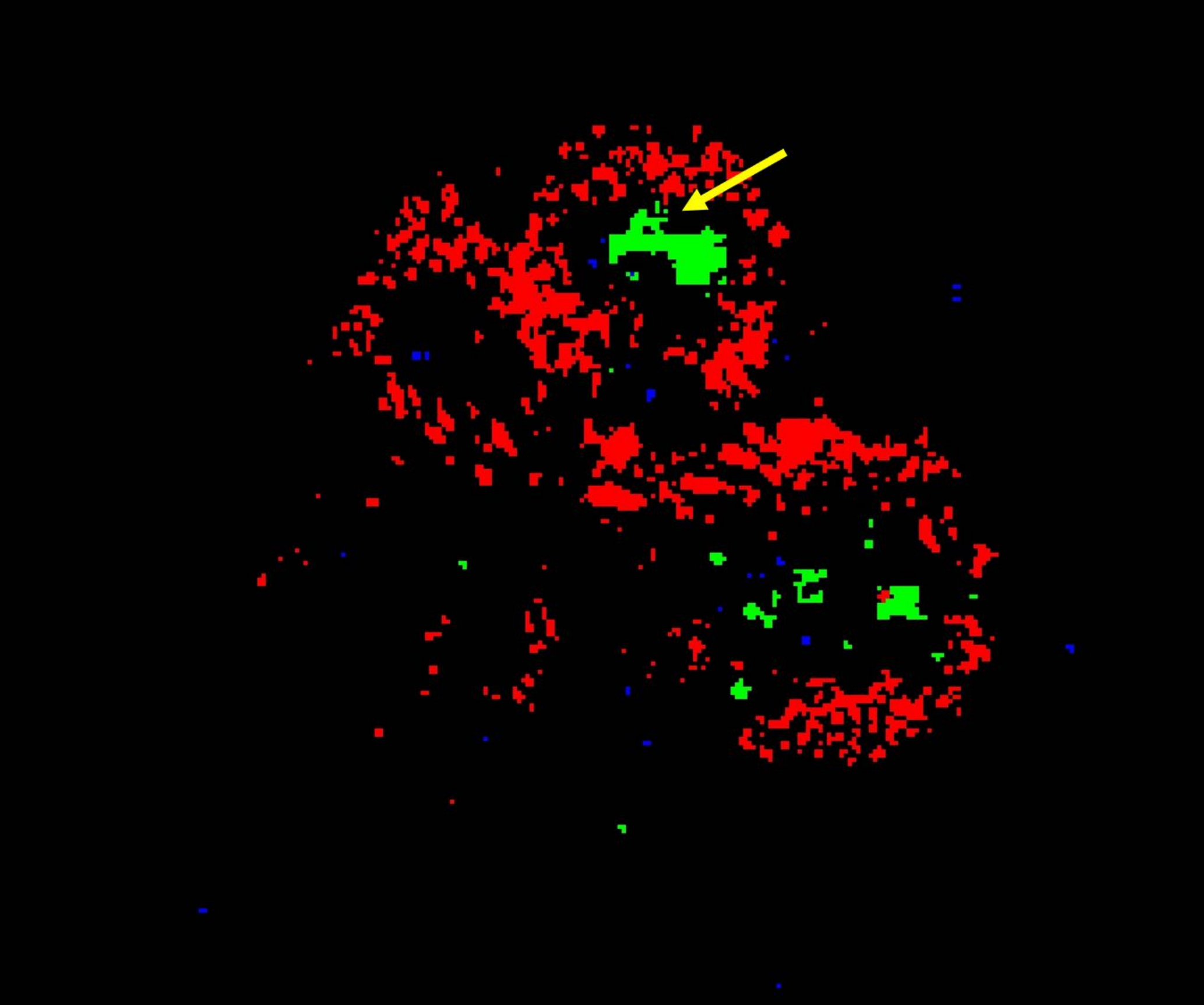}}
	\subfloat[Ours]{\includegraphics[width=1.69in]{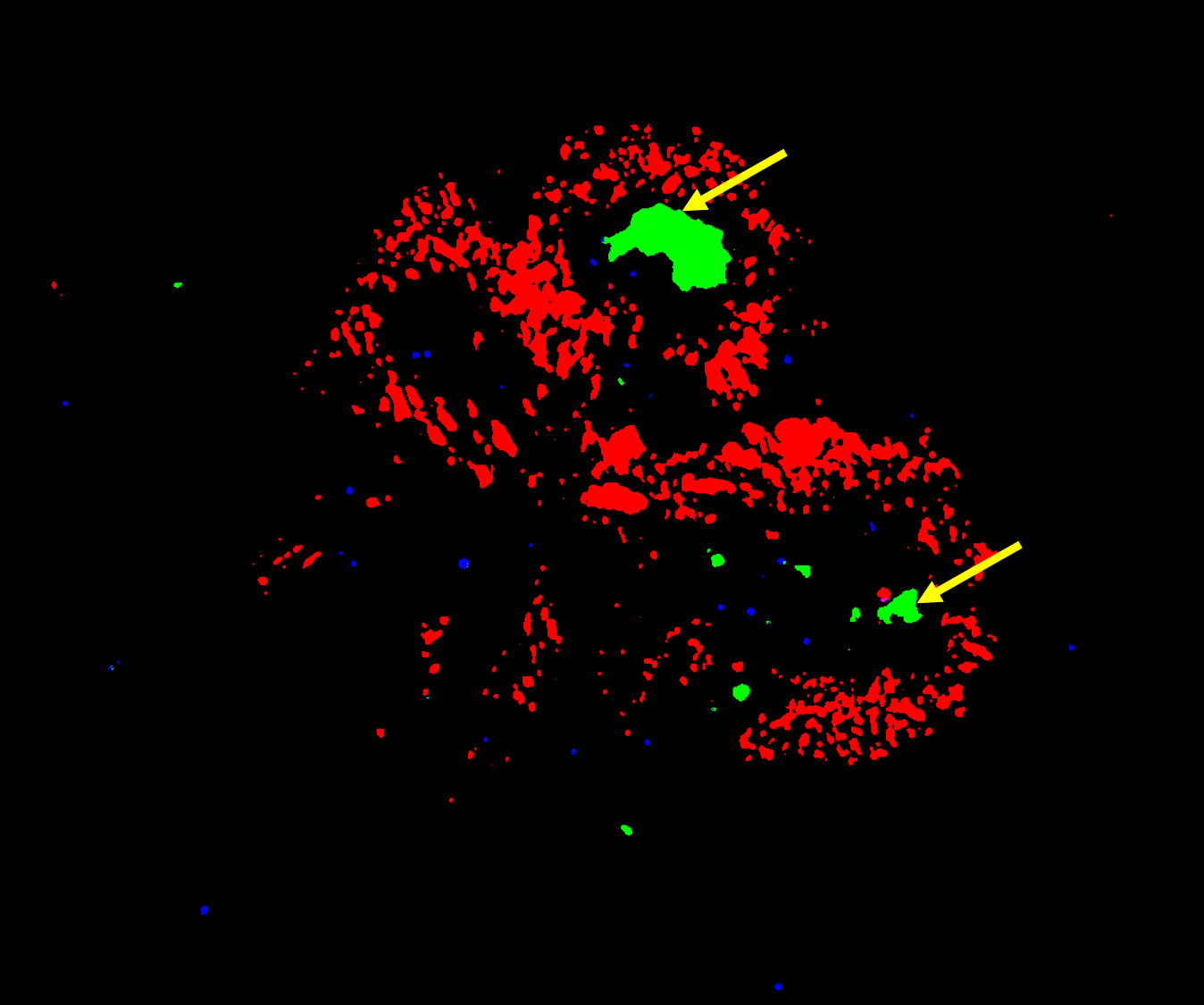}}
	
	\caption{Qualitative results of our proposed PMCNet and other state-of-the-art methods, including UNet, ResUNet, DenseUNet, UNet++ and DeepLabv3+.  Compared with other methods, our network can produce a more complete and reliable lesion map. Red, green, blue, and pink markings denote EX, HE, MA, and SE, respectively. The yellow arrows indicate the differences between the predictions of these methods and the ground truth.}
	\label{fig_6}
\end{figure*}

\textbf{Analysis of DAB:}
In the PFF block, we indiscriminately fused all the features from three adjacent scale series. Although this strategy can provide multi-scale information progressively, it leads to the inconsistency of features, because there exists an intrinsic difference in multi-scale features; thus, the segmentation performance decreased in MA compared to the baseline. Thus, it is necessary to dynamically select the fused features, and this strategy will suppress noises and improve the feature consistency among different scales during training. Therefore, we used DAB to highlight useful features and reduce feature inconsistency for fundus lesion segmentation.

\textbf{First}, compared with baseline+PFF(DWConv), it is clear from the results that the segmentation performance of MA can be improved significantly after adopting DAB, which demonstrates that the proposed DAB will reduce the feature conflict and inconsistency. The performance of HE decreases slightly, and for DAB, it performs dynamic attention for small lesions, and HE have various sizes, while  MA are isolated small dots, during the feature selection, some useful information around noise for HE may be filtered out, because it is a relatively continuous region rather than an isolated dots, and thus it hurts HE. Compared with baseline, the mAUC and AUC score of each type of lesion can be improved, which shows that the proposed PFF and DAB will make full use of multi-scale fundus lesion features and reduce the feature gap among different scales. \textbf{Second}, we also did the experiment with only add DAB to the baseline, and the results show that DAB can improve the segmentation results of fundus lesions, especially for MA, which indicates that DAB can solve the feature inconsistency problem of multi-scale feature fusion well. 

We also compared DAB with other attention blocks to demonstrate the effectiveness of DAB. For a fair comparison, we maintained the same model parameters for each attention block. The results in Table \ref{table_1} show that the proposed DAB achieves the best fundus lesion segmentation performance on the IDRiD dataset compared with SE \cite{hu2018squeeze} and CBAM \cite{woo2018cbam} attention blocks, demonstrating the effectiveness of the proposed block. For the SE block, although it can assign different attention weights to each feature channel to explore different feature patterns of fundus lesions, it ignores spatial pixel-wise relationships, resulting in poor performance in such pixel-level lesion segmentation tasks. Although CBAM blocks apply channel attention and spatial attention, they are applied consecutively and not in a parallel manner, which causes ambiguity in channel attention and spatial attention. Therefore, the proposed DAB considers both the channel and spatial attention in a parallel branch for the input feature maps. In addition, we also apply point-wise attention because each feature point has its own attention weights by dynamic learning. By aggregating triplet attention, DAB improved the learning ability of the model and performed better than SE and CBAM.
\begin{table*} [ht]
	\caption{The fundus lesion segmentation results of the proposed PFF and DAB blocks combined with other CNN backbones on IDRiD dataset. 'Param' denotes the parameters of the network, and the test time is evaluated on an NVIDIA GTX 1080Ti GPU with the input resolution of 960$\times$1440.}
	\center
	\setlength{\tabcolsep}{3pt}
	\renewcommand\arraystretch{1.2}
	\begin{tabular}{p{120pt}|p{23pt}p{23pt}p{23pt}p{23pt}|p{23pt}p{23pt}p{23pt}|p{36pt}|p{41pt}}
		\hline
		Method               &EX      &HE      &MA      &SE      &mAUC    & Dice& IoU& Param& Test time\\
		\hline
		ResNet50 \cite{he2016deep}           & 71.13&31.87&32.00&\textbf{53.10}&47.03&44.52&32.99&25.26M&0.1851s\\
		ResNet50+PFF+DAB                  & \textbf{79.33}&\textbf{50.06}&\textbf{38.00}&42.86&\textbf{52.56}&\textbf{47.12}&\textbf{35.58}&30.42M&0.2247s\\
		\hline
		VGG16 \cite{simonyan2014very}       & 75.68&59.86&42.26&65.69&60.87&49.98&39.54&22.47M&0.2987s\\
		VGG16++PFF+DAB            &\textbf{82.38}&\textbf{64.86}&\textbf{44.39}&\textbf{69.58}&\textbf{65.30}&\textbf{52.21}&\textbf{40.92}&25.63M&0.3242s\\
		\hline
		EfficientnetB0  \cite{tan2019efficientnet}       & 81.95&\textbf{69.73}&41.90&66.33&64.98&51.69&39.42&6.75M&0.1481s\\
		EfficientnetB0++PFF+DAB    &\textbf{87.24}&67.05&\textbf{46.94}&\textbf{71.11}&\textbf{68.08}&\textbf{56.02}&\textbf{43.12}&8.57M&0.1851s\\
		\hline
	\end{tabular}
	\label{table_2}
\end{table*}

\textbf{Generalization to Different Backbones:}
To demonstrate the generalization of the two proposed blocks, we adopted other CNN backbones that are widely used for computer vision tasks, including VGG16 \cite{simonyan2014very}, ResNet50 \cite{he2016deep}, and EfficientnetB0 \cite{tan2019efficientnet}. We combined the proposed PFF and DAB blocks with these backbones for fundus lesion segmentation, and the results are presented in Table \ref{table_2}, we can see from the results that the baseline models integrate with the proposed blocks can achieve significant performance improvement in AUC, Dice and IoU, which indicates that the proposed blocks has good generalization ability to different backbones. 
From the results, we can see different backbone network can make a significant difference, although ResNet is a robust network in natural image processing, it does not well adapt to the segmentation of small lesions.
We also reported the parameters of the model and the inference speed, it can be seen that the increased number of parameters is acceptable compared with baseline, and the test time has not been particularly affected.  Some visual examples of the lesion segmentation results using different backbones are shown in Fig. \ref{fig_5}. From the segmentation results, we can see that the proposed module can produce more reliable lesion segmentation maps. In terms of the lesion segmentation results, the EfficientnetB0 based model achieves the best result; therefore, we use it 
as the backbone, when compared with other state-of-the-art methods.

\begin{figure*}[ht]
	\centering
	\subfloat{\includegraphics[width=1.60in]{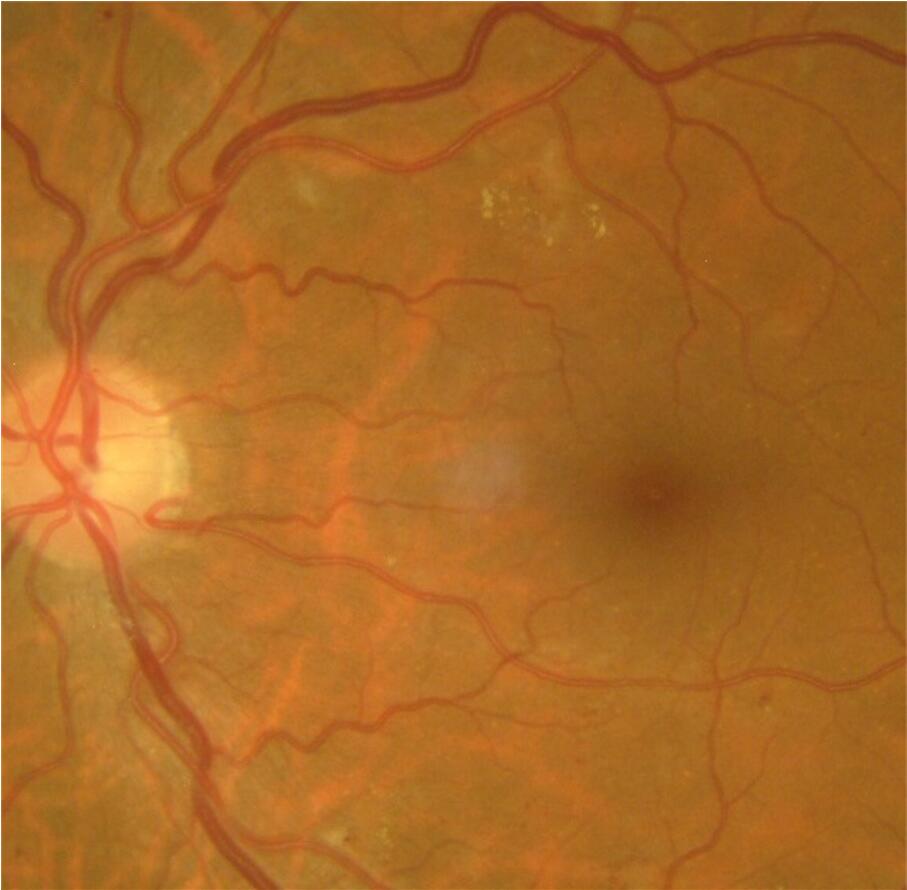}}
	\subfloat{\includegraphics[width=1.60in]{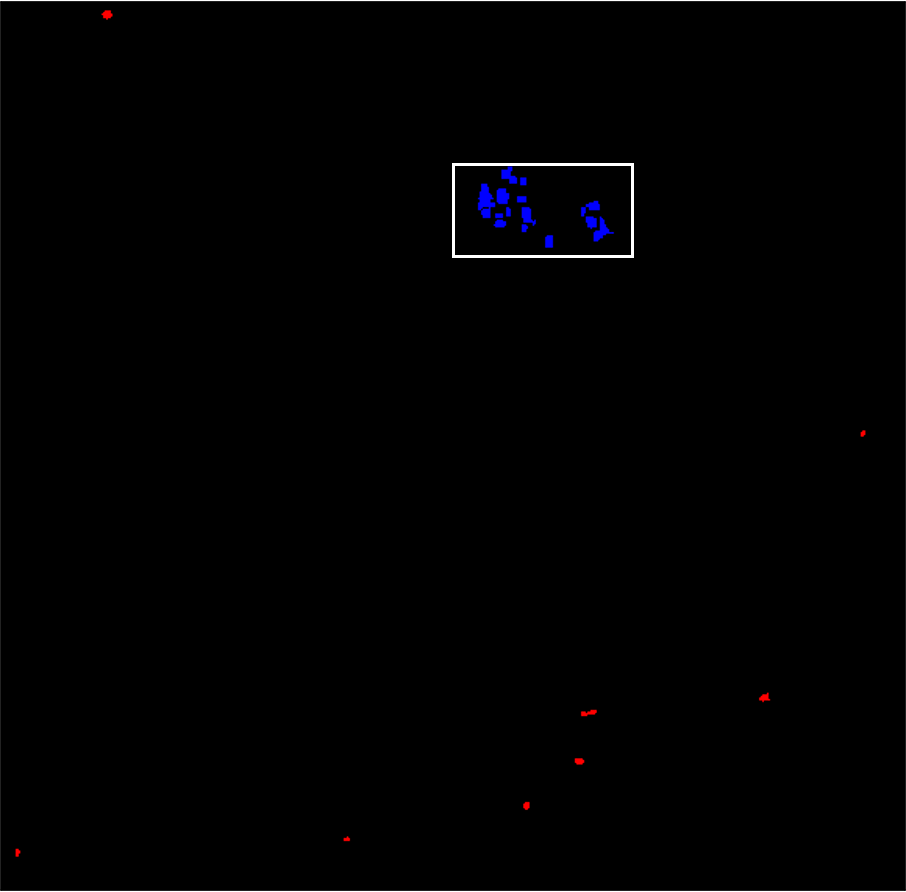}}
	\subfloat{\includegraphics[width=1.60in]{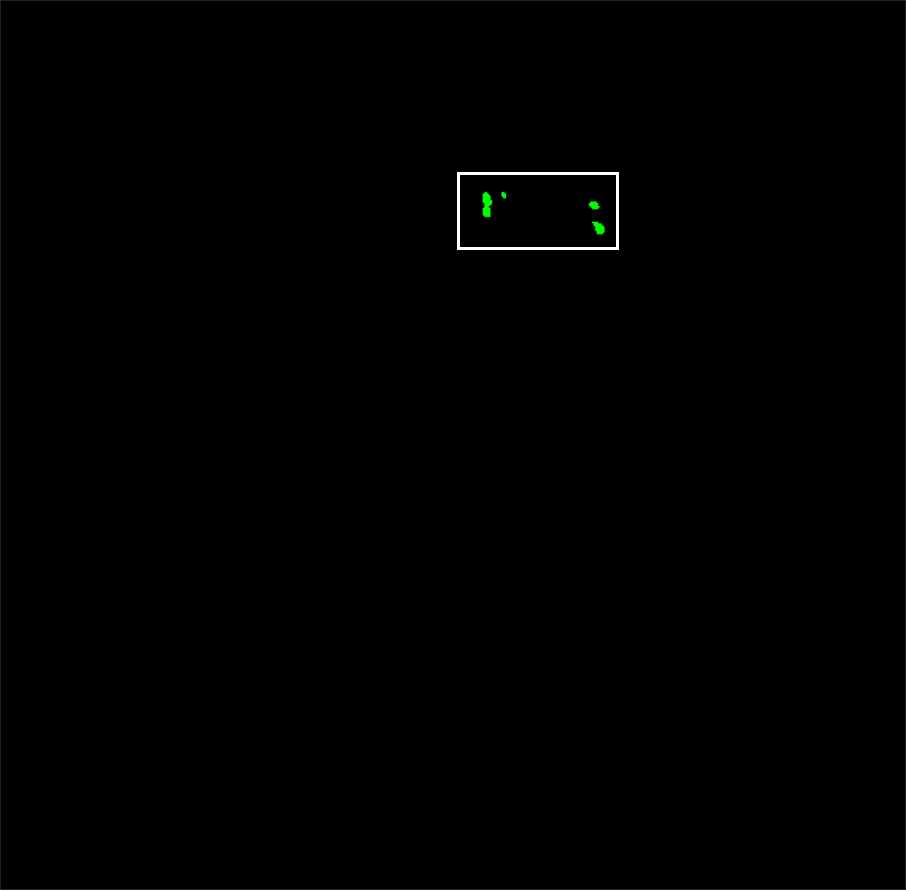}}
	\subfloat{\includegraphics[width=1.60in]{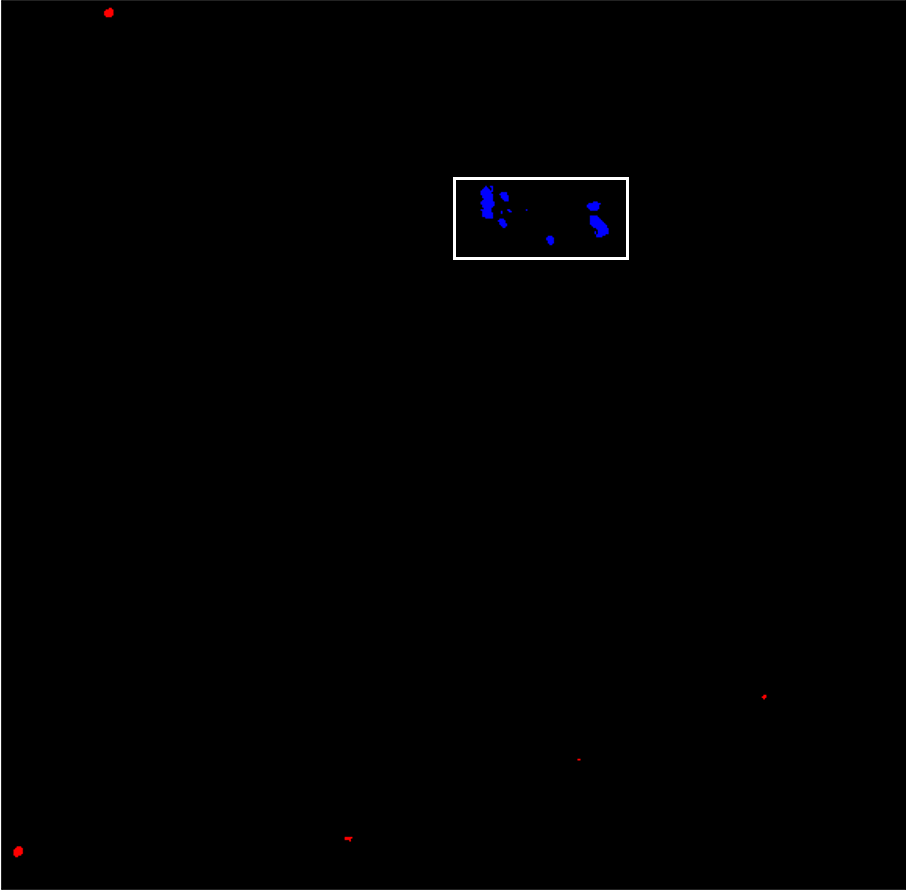}}

	\subfloat{\includegraphics[width=1.60in]{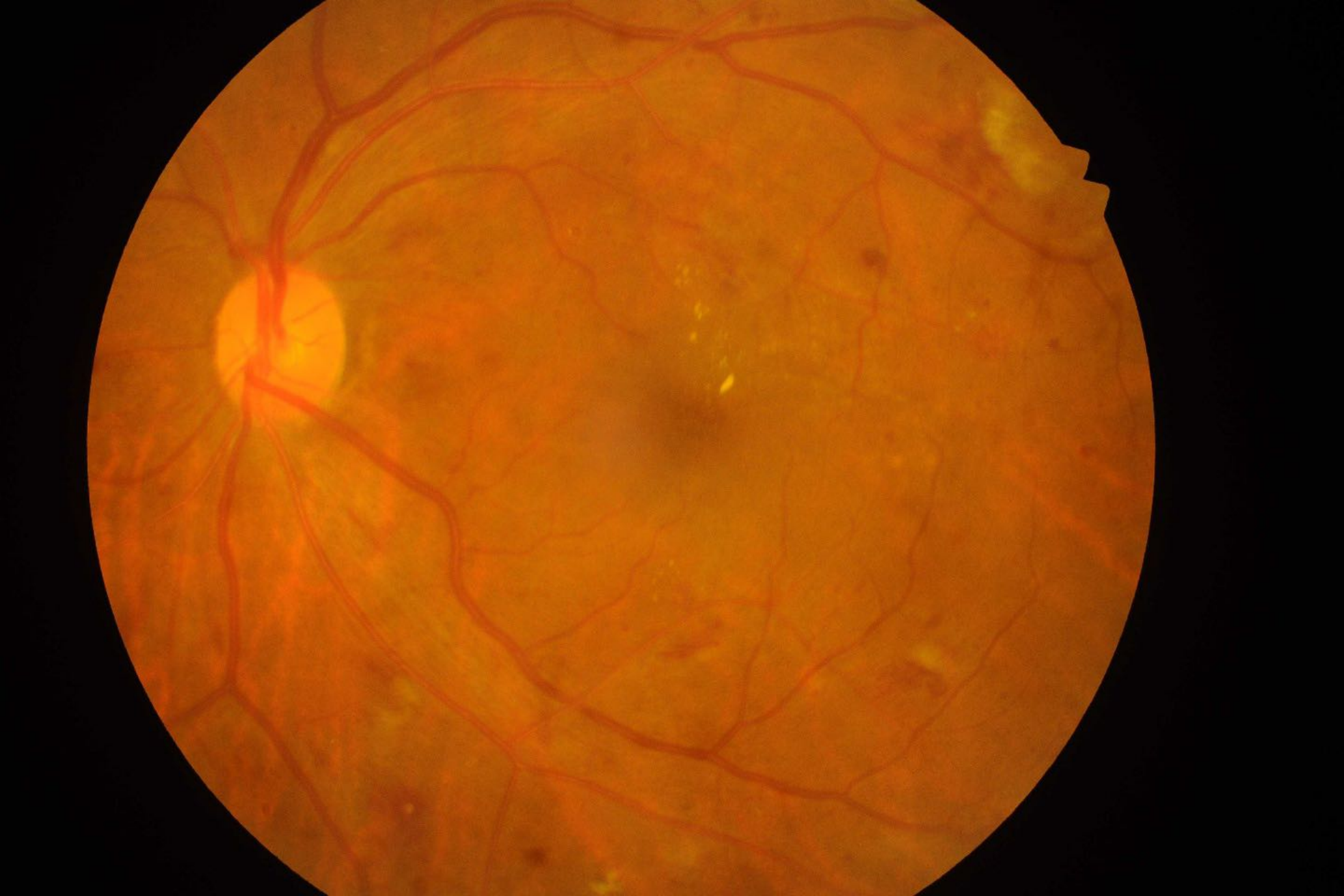}}
	\subfloat{\includegraphics[width=1.60in]{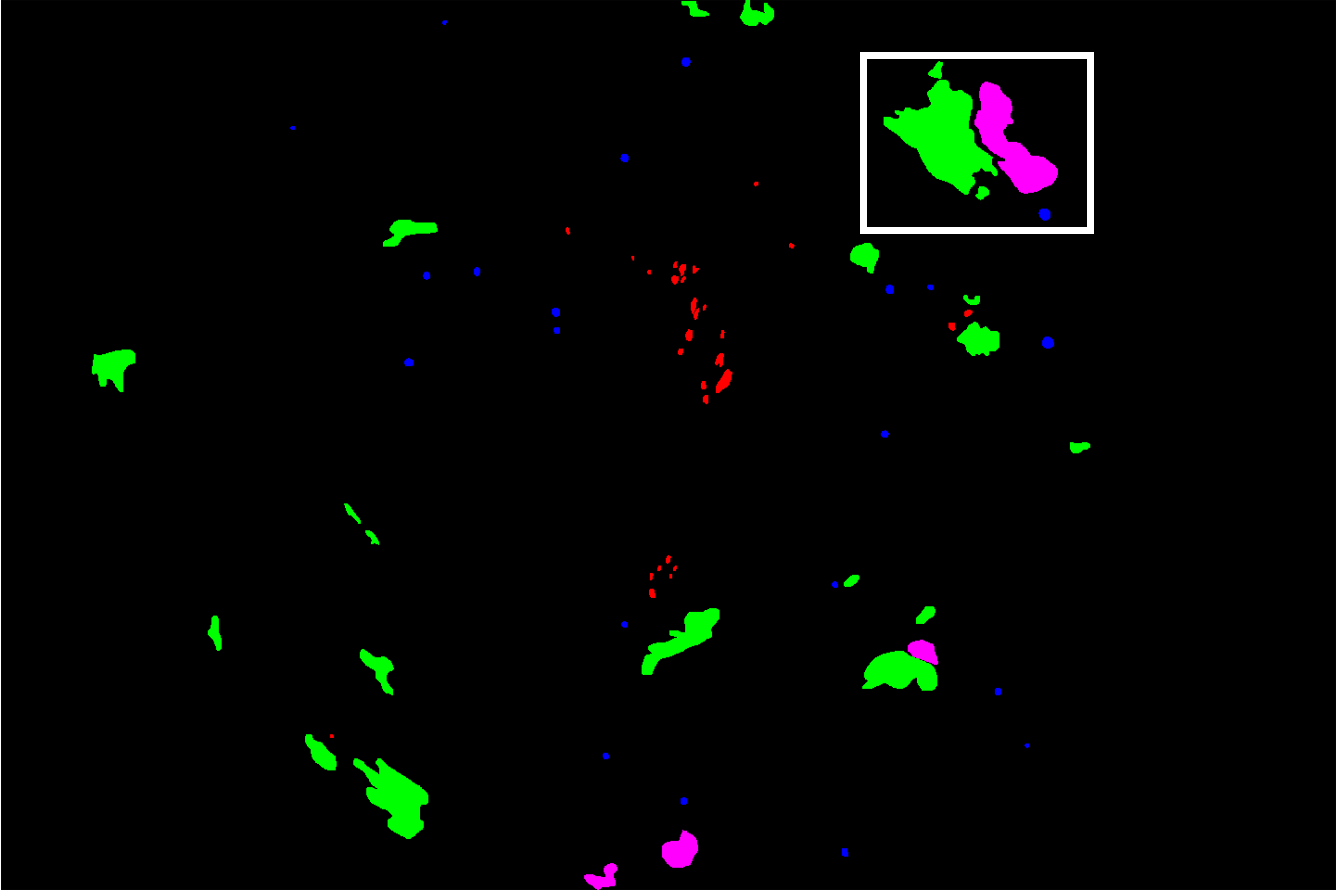}}
	\subfloat{\includegraphics[width=1.60in]{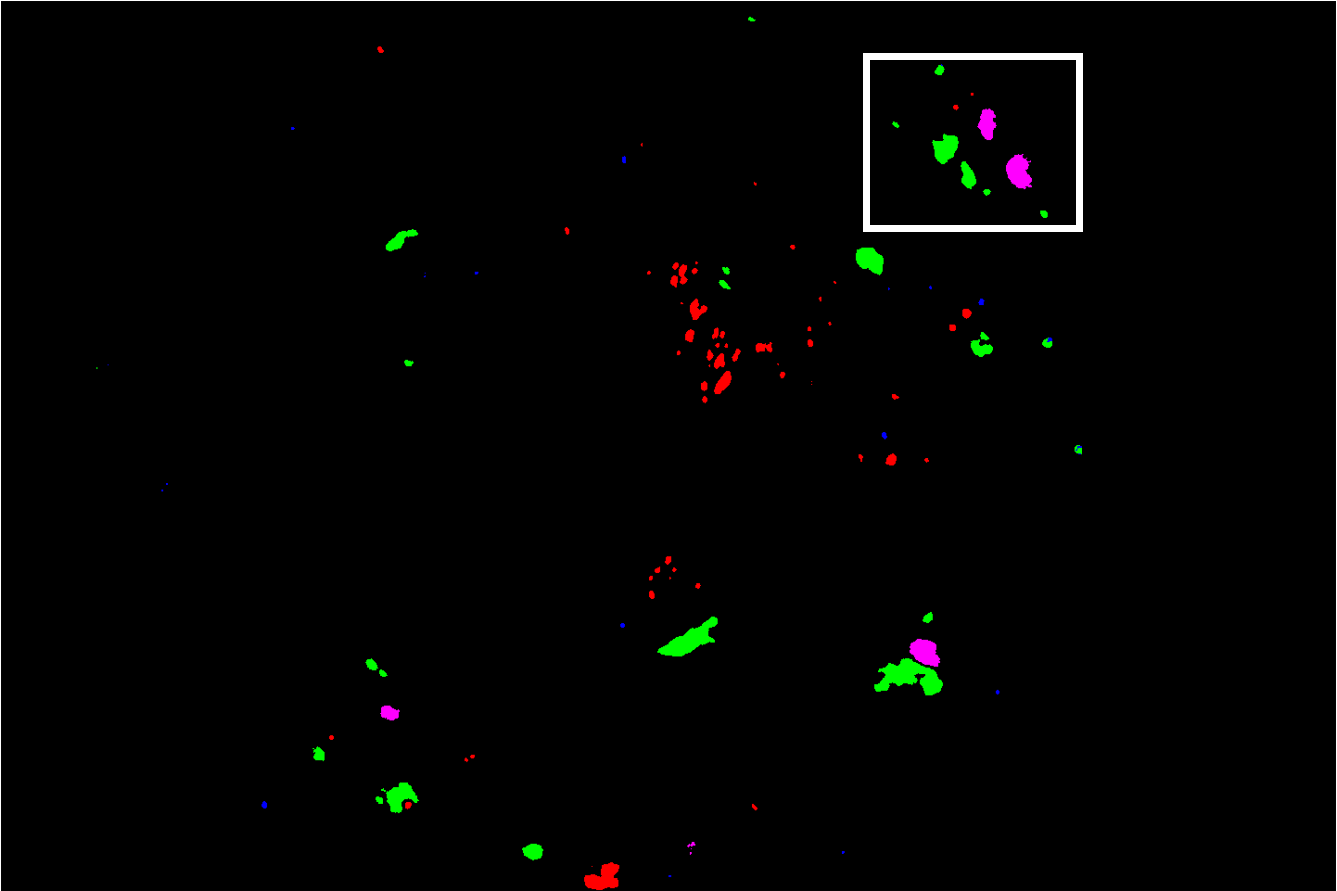}}
	\subfloat{\includegraphics[width=1.60in]{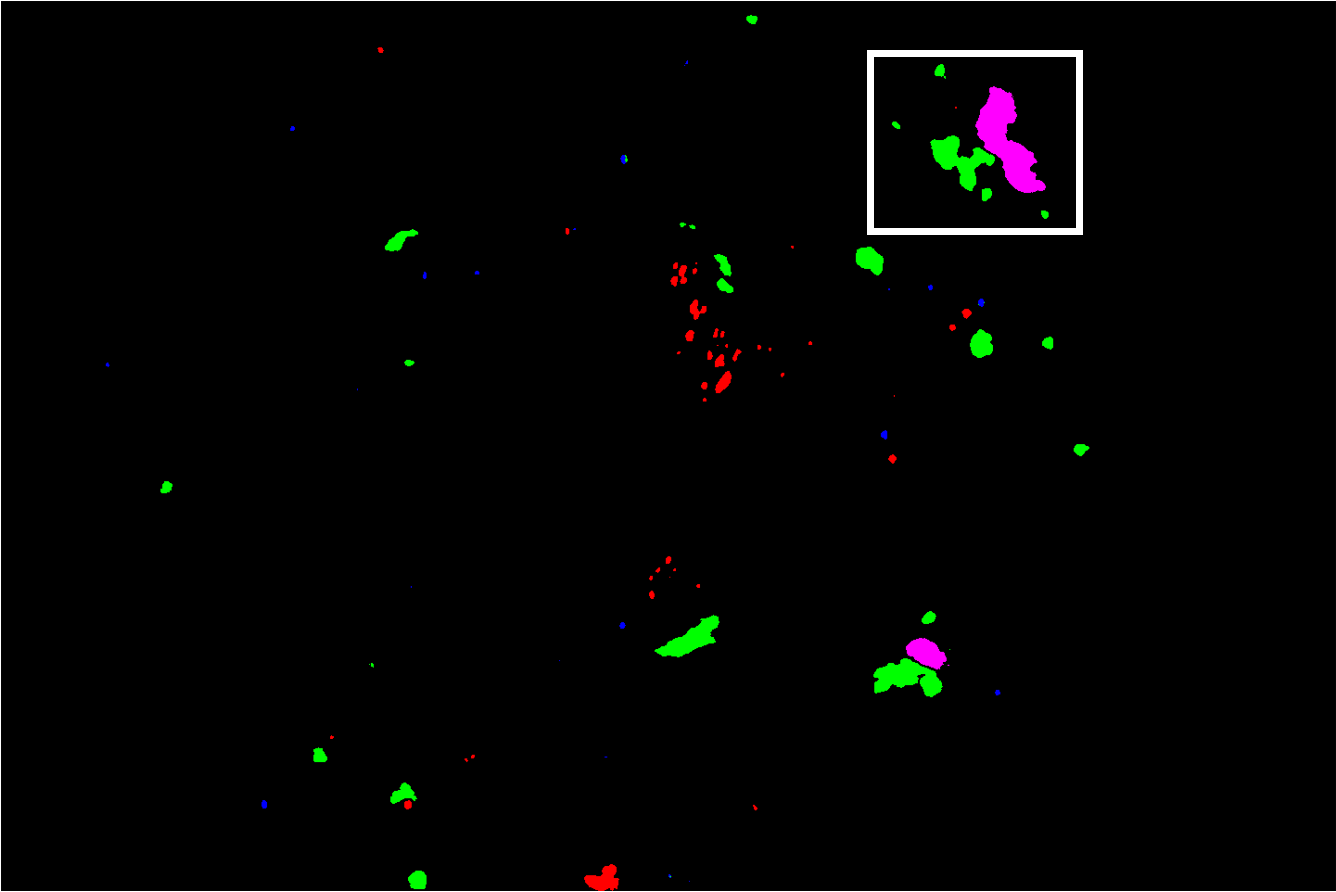}}
	\setcounter{subfigure}{0}
	
	\subfloat[fundus]  {\includegraphics[width=1.60in]{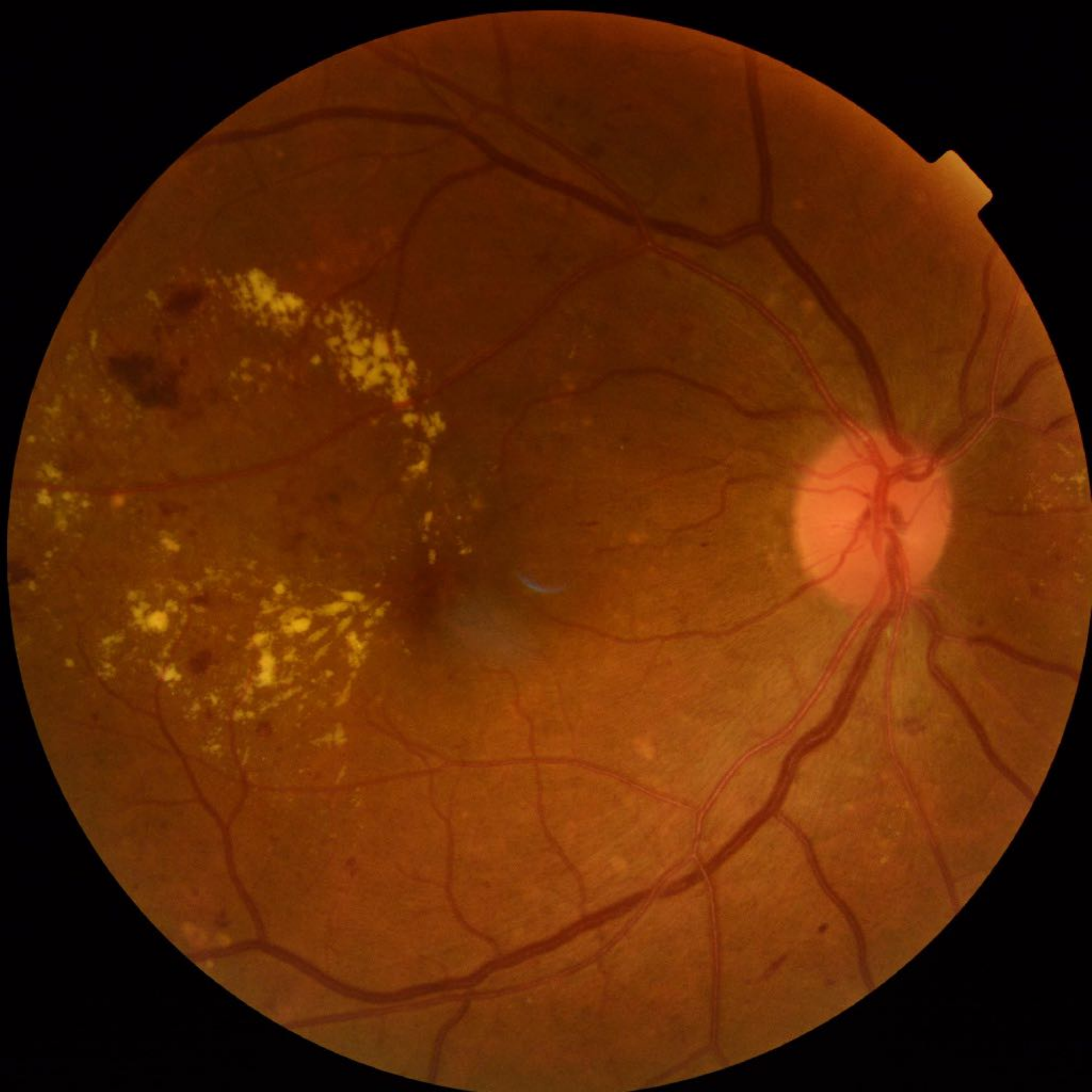}}
	\subfloat[GT]      {\includegraphics[width=1.60in]{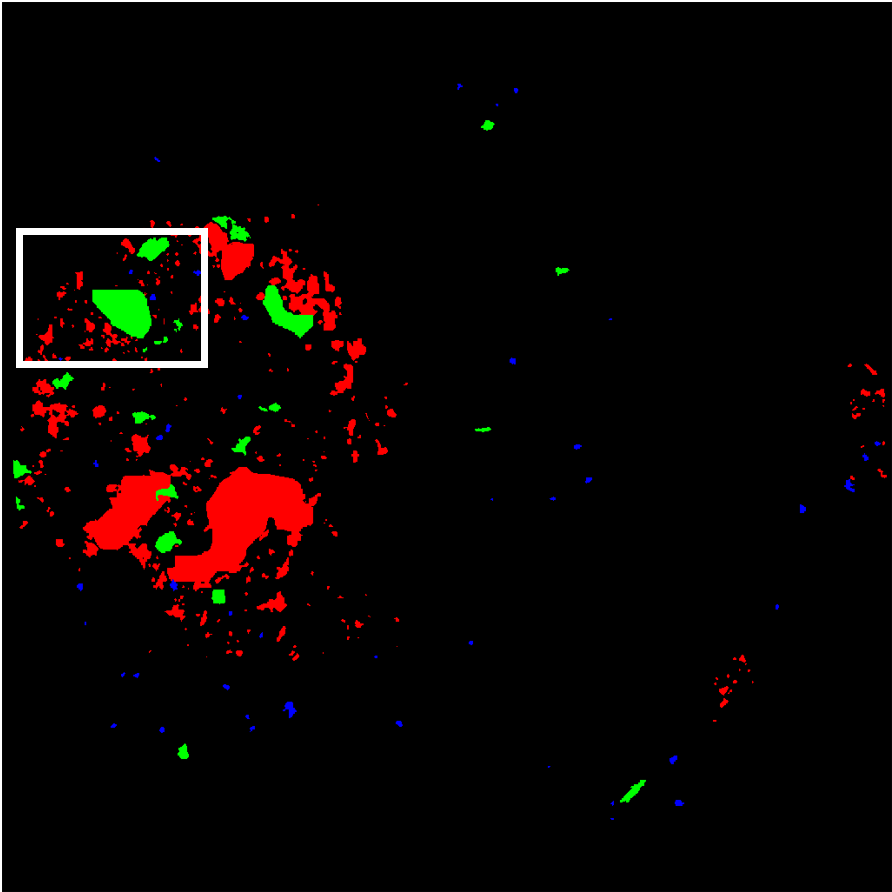}}
	\subfloat[Baseline]{\includegraphics[width=1.60in]{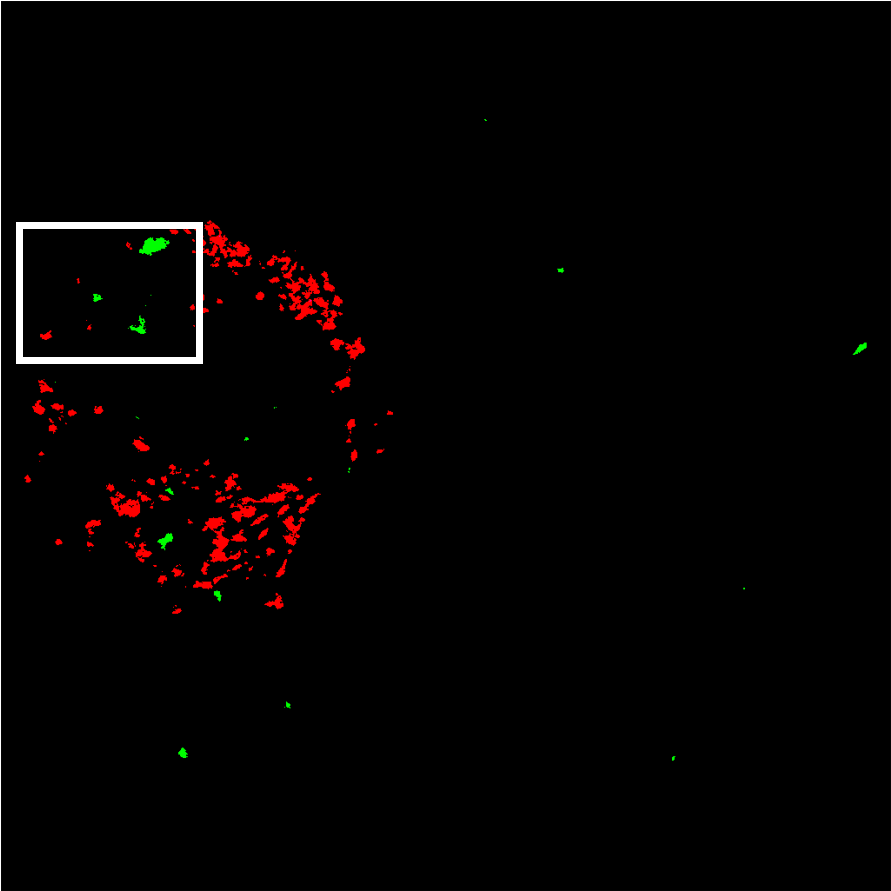}}
	\subfloat[Ours]    {\includegraphics[width=1.60in]{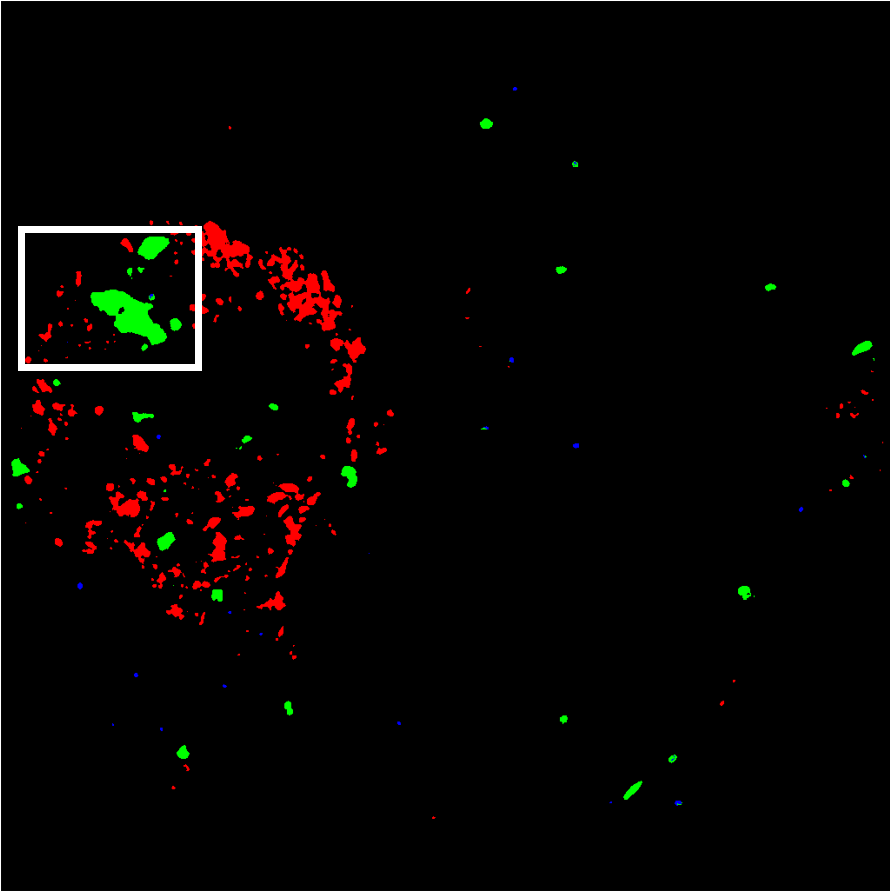}}
	
	\caption{Three examples from E-ophtha, IDRiD and DDR  datasets from top to bottom, respectively. Since the lesions in E-ophtha dataset are too small to be visualized, we cropped the lesion region to better show the lesion information. There are significant differences in resolution, color and lesion distribution among the three datasets. Red, green, blue, and pink markings denote EX, HE, MA, and SE, respectively.}
	\label{fig_7}
\end{figure*}
\begin{figure*}[htb]
	\centering
	\subfloat[fundus]{\includegraphics[width=2.2in]{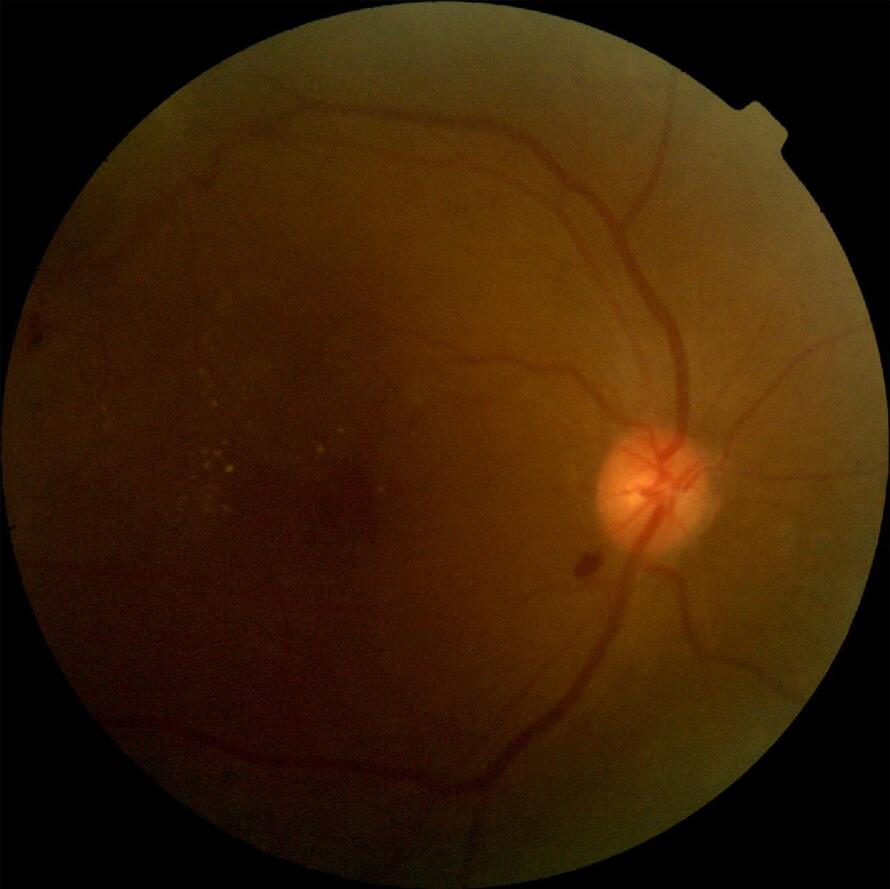}}
	\subfloat[GT]{\includegraphics[width=2.2in]{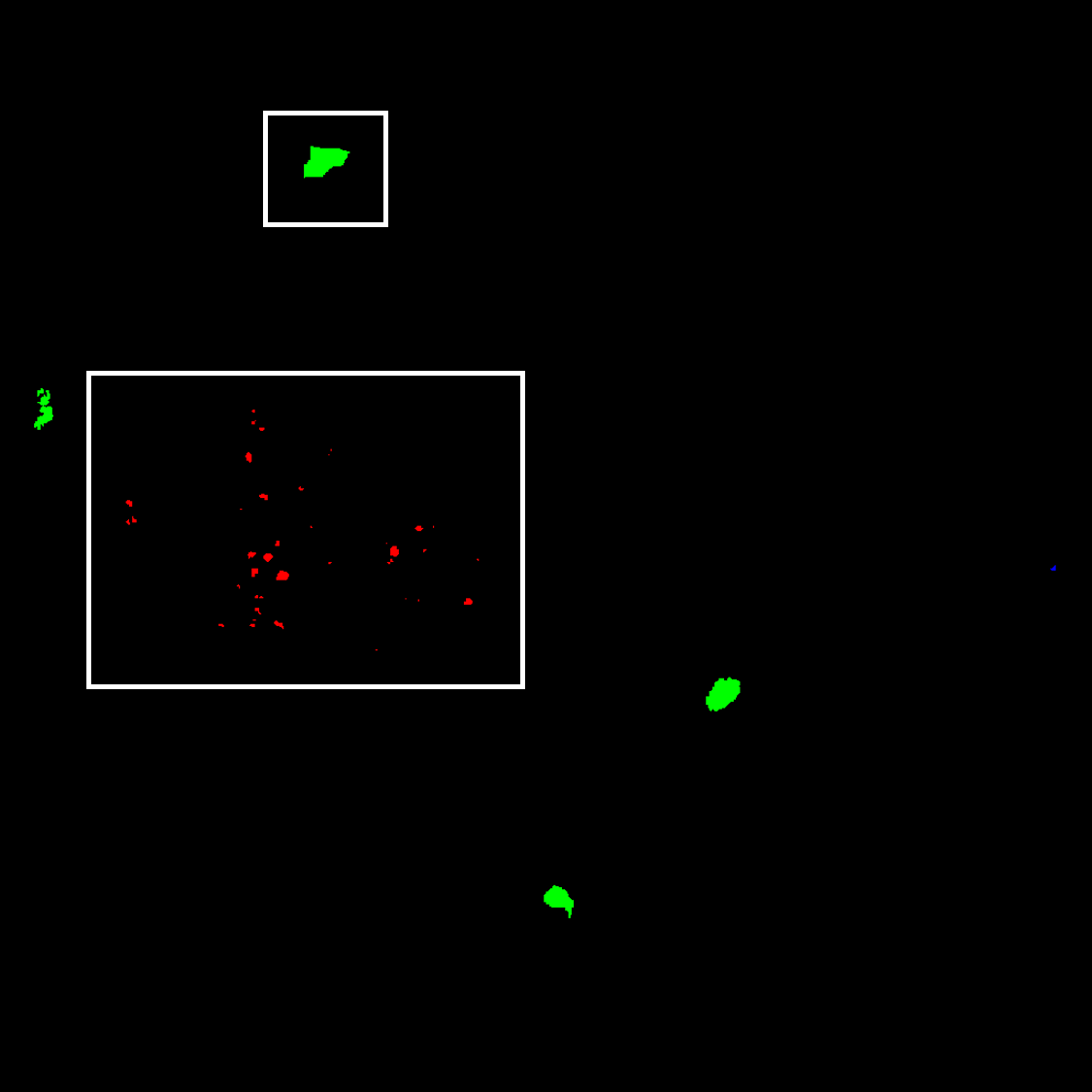}}
	\subfloat[Ours]{\includegraphics[width=2.2in]{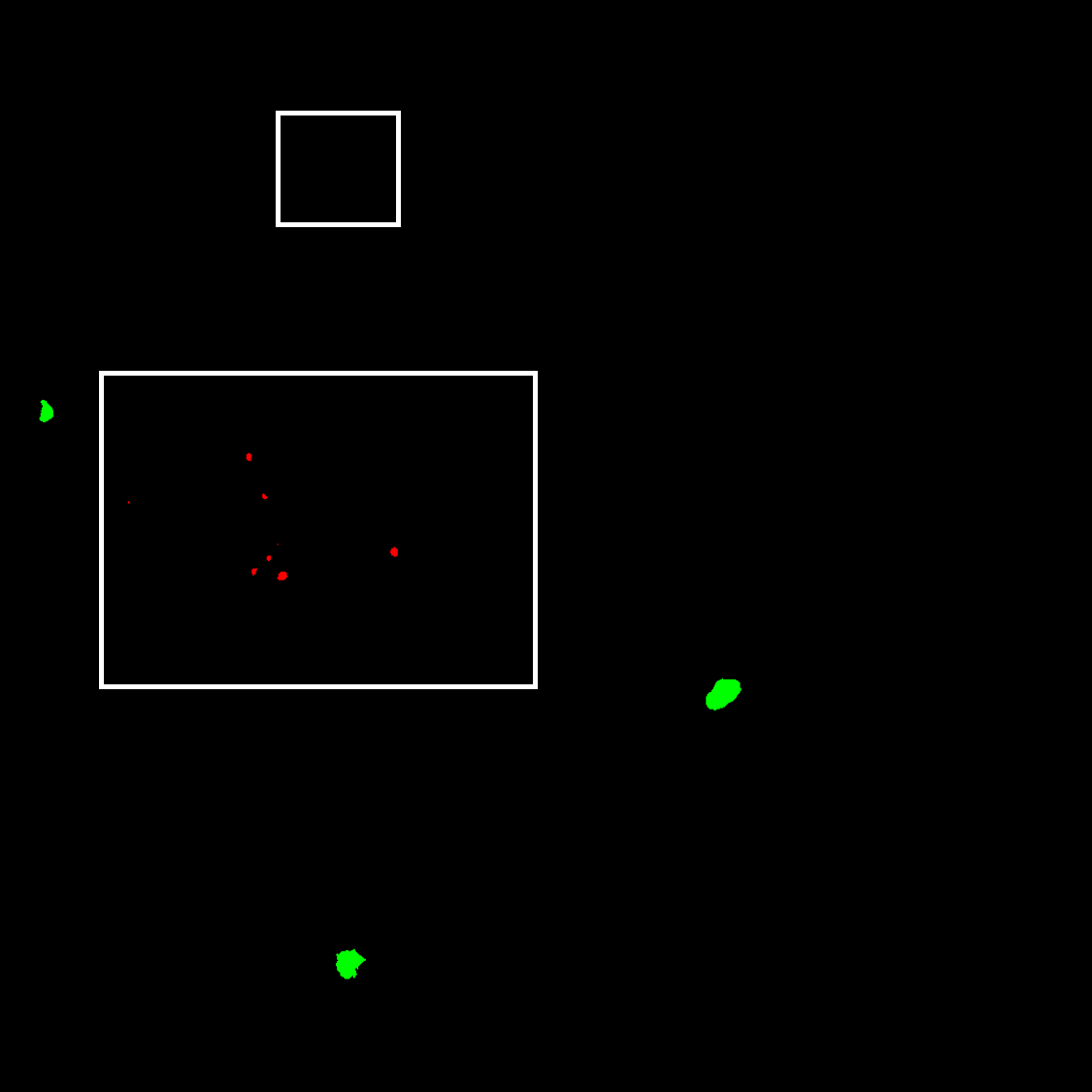}}
	
	\caption{ A failure case from DDR dataset with low contrast and small lesions. (a) Original fundus image. (b) The ground truth. (c) The segmentation results of our method. It can be seen that due to the low contrast, some small lesions are missed in the segmentation result.}
	\label{fig_8}
\end{figure*}

\subsection{Comparisons with Other State-of-the-art Methods}

To further measure the multi-class lesion segmentation performance of our proposed method, we compared it with eleven state-of-the-art methods on the three benchmark datasets, which can be grouped into UNet-based networks and non-UNet-based networks. UNet-based networks include UNet \cite{ronneberger2015u}, UNet++ \cite{zhou2018unet++}, ResUNet \cite{zhang2018road},  Att-UNet \cite{oktay2018attention} and DenseUNet \cite{li2018h}. Non-UNet-based networks include DeepLabv3+ \cite{chen2018encoder}, PSPNet \cite{zhao2017pyramid}, HED  \cite{xie2015holistically},  FCRN \cite{mo2018exudate},  CASENet \cite{yu2017casenet}, and L-Seg \cite{guo2019seg}.
Table \ref{table_idrid}, \ref{table_ddr} and \ref{table_e-ophtha} lists the quantitative results for the three datasets, including the AUC score of each lesion, mean AUC, Dice and IoU scores. The results for UNet, ResUnet, DenseUnet, UNet++, DeepLabv3+ and PSPNet were implemented by us, and other results were directly taken from paper \cite{guo2019seg}. To make a fair comparison, for UNet, ResUNet, UNet++, Att-UNet, DeepLabv3+ and PSPNet, we adopt EfficientnetB0 as the backbone, for DenseUNet, we adopt DenseNet as the backbone.

We can see that our method achieves superior performance over these compared methods on three datasets, especially on the DDR dataset, which is the largest and most challenging dataset. The proposed method obtains more complete lesion maps through PFF and DAB, which demonstrates the effectiveness of our proposed method.
We also display some qualitative results for some representative state-of-the-art models, as shown in Fig. \ref{fig_6}.
We can see from the results that UNet and its variants produce more background noises, especially for UNet, ResUNet, and UNet++.
This is because they do not consider the multi-scale feature interaction and feature consistency when performing feature fusion, which leads to inconsistent  features; thus, some backgrounds are regarded as lesions.
Note that DenseUNet can achieve satisfactory results owing to the powerful feature extraction ability of dense connections.
For DeepLabv3+, even though it produces satisfactory results in natural images, it generates a blurry and coarse lesion segmentation map (green markings), and details are missing, which is caused by its direct up-sampling to restore the lesion information.
Our method can well handle these issues and produce more accurate results. The segmentation results from the three datasets are shown in Fig. \ref{fig_7}. The three datasets differ greatly in resolution, lesion distribution and image contrast, especially for the E-ophtha dataset, the lesions are small, making it difficult to segment. Since the lesions in E-ophtha dataset are too small to be visualized, and thus we cropped the lesion region to better show the lesion information. Certainly, there are the limitations of our proposed method when processing fundus images of poor quality, some lesions may be missed, such as fundus images with low contrast, as shown in Fig. \ref{fig_8}, some HE and EX lesions are missed due to the poor image quality.

\begin{table}[h!] \scriptsize
	\center
	\caption{Comparison results of different fundus  lesion segmentation methods on IDRiD dataset.}
	\setlength{\tabcolsep}{2pt}
	\renewcommand\arraystretch{1.2}
	\begin{tabular}{p{57pt}|p{20pt}p{20pt}p{20pt}p{20pt}|p{22pt}p{20pt}p{20pt}p{20pt}}
		\hline
		
		Methods& EX     &HE     &MA     &SE &mAUC    &Dice &IoU &Param  \\
		\hline
		UNet \cite{ronneberger2015u}   &65.94&14.08&28.01&56.32&41.09&41.72&32.07& 7.86M  \\ 
		ResUNet \cite{zhang2018road}   &62.82&25.68&25.65&33.67&36.96&33.14&23.57 &7.25M  \\ 
		DenseUNet \cite{li2018h}       &80.54&41.53&38.64&64.68&56.35&49.82&37.71 & 6.58M \\
		UNet++ \cite{zhou2018unet++}   &78.18&12.88&21.74&43.76&39.14 &34.93&25.08 & 7.95M\\
		Att-UNet \cite{oktay2018attention}   &65.29&17.40&28.32&56.02&41.75&42.39&33.84& 7.96M   \\
		\hline
		DeepLabv3+ \cite{chen2018encoder}&77.62 &61.57 &22.92 &64.03 &56.53&48.47&37.44 &7.67M  \\
		PSPNet \cite{zhao2017pyramid}    &51.31 &21.89 &17.99 &43.53 &33.69&27.19&21.56 &7.89M \\
		HED  \cite{xie2015holistically}  &77.23 &50.84 &42.71 &66.73 &59.38&-&- &- \\
		FCRN \cite{mo2018exudate}        &54.72 &42.00 &33.83 &51.54 &45.52&-& -& -\\
		CASENet \cite{yu2017casenet}     &75.64 &44.62 &39.92 &32.75 &48.23&-& -&- \\
		L-Seg \cite{guo2019seg}          &79.45 &63.74 &46.27 &\textbf{71.13} &65.15&-&-&-\\
		\hline
		PMCNet(Ours)           &\textbf{87.24}&\textbf{67.05}&\textbf{46.94}&71.11&\textbf{68.08}&\textbf{56.02}&\textbf{43.12}&8.57M \\
		\hline
	\end{tabular}
	\label{table_idrid}
\end{table}

\begin{table}[h!] \scriptsize
	\center
	\caption{Comparison results of different fundus  lesion segmentation methods on DDR dataset.}
	\setlength{\tabcolsep}{2pt}
	\renewcommand\arraystretch{1.2}
	\begin{tabular}{p{57pt}|p{20pt}p{20pt}p{20pt}p{20pt}|p{22pt}p{20pt}p{20pt}p{20pt}}
		\hline
		
		Methods& EX     &HE     &MA     &SE &mAUC    &Dice &IoU &Param  \\
		\hline
		UNet \cite{ronneberger2015u}   &45.86&9.70&9.55&10.94&19.01&28.23&26.37& 7.86M  \\ 
		ResUNet \cite{zhang2018road}   &40.42&15.32&9.04&8.69&18.37&27.55&25.73 &7.25M  \\ 
		DenseUNet \cite{li2018h}       &54.59&37.32&18.04&24.34&33.58&37.29&30.19 & 6.58M \\
		UNet++ \cite{zhou2018unet++}   &45.27&12.05&11.54&7.94&19.20&31.47& 27.36& 7.95M\\
		Att-UNet \cite{oktay2018attention}   &46.26&11.30&8.94&12.73&19.80& 31.96&27.93& 7.96M   \\
		\hline
		DeepLabv3+ \cite{chen2018encoder} &54.63 &34.20 &3.04 &29.10 &30.24 &36.42& 29.83 &7.67M  \\
		PSPNet \cite{zhao2017pyramid}     &34.97 &21.18 &5.57 &9.09 &17.71 & 26.83&24.84 &7.89M \\
		HED  \cite{xie2015holistically}   &42.41 &19.19 &4.96 &9.72 &19.07&-&- &- \\
		FCRN \cite{mo2018exudate}         &19.96 &12.55 &1.34 &4.54 &9.60&-&-& -\\
		CASENet \cite{yu2017casenet}      &27.77 &26.25 &10.05 &13.04 &19.28&-&-&- \\
		L-Seg \cite{guo2019seg}           &\textbf{55.46} &35.86 &10.52 &26.48 &32.08&-&-&-\\
		\hline
		PMCNet(Ours)           &54.30&\textbf{39.87}&\textbf{19.94}&\textbf{31.64}&\textbf{36.44}&\textbf{39.31}&\textbf{32.29} &8.57M \\
		\hline
	\end{tabular}
	\label{table_ddr}
\end{table}

\begin{table}[h!] \scriptsize
	\center
	\caption{Comparison results of different fundus  lesion segmentation methods on E-ophtha dataset.}
	\setlength{\tabcolsep}{3pt}
	\renewcommand\arraystretch{1.2}
	\begin{tabular}{p{57pt}|p{20pt}p{20pt}|p{22pt}p{20pt}p{20pt}p{23pt}}
		\hline
		
		Methods& EX    &MA    &mAUC    &Dice &IoU &Param  \\
		\hline
		UNet \cite{ronneberger2015u}   &28.54 &14.85 &21.69&15.65&12.63& 7.86M  \\ 
		ResUNet \cite{zhang2018road}   &35.48 &12.50 &23.99&16.53&14.34&7.25M  \\ 
		DenseUNet \cite{li2018h}       &38.62 &18.27 &28.44&31.56&20.22& 6.58M \\
		UNet++ \cite{zhou2018unet++}   &34.38 &16.42 &25.39&17.41&18.32& 7.95M\\
		Att-UNet \cite{oktay2018attention}   &28.94 &15.16 &22.05&16.16&13.21& 7.96M   \\
		\hline
		DeepLabv3+ \cite{chen2018encoder}&\textbf{53.24} &4.65 &28.95&29.32&21.67&7.67M  \\
		PSPNet \cite{zhao2017pyramid}    &29.03 &10.21 &19.62&14.83&11.63 &7.89M \\
		HED  \cite{xie2015holistically}  &13.45 &6.39  &9.92&-&- &- \\
		FCRN \cite{mo2018exudate}        &16.75 &2.69  &9.72&-&-& -\\
		CASENet \cite{yu2017casenet}     &17.15 &15.65 &16.40&-&-&- \\
		L-Seg \cite{guo2019seg}          &41.71 &16.87 &29.29&-&-&-\\
		\hline
		PMCNet(Ours)                     &51.20 &\textbf{30.60} &\textbf{40.90}&\textbf{45.43}&\textbf{30.70}&8.57M \\
		\hline
	\end{tabular}
	\label{table_e-ophtha}
\end{table}

\section{Conclusion and Future Work}
\label{Conclusion}
In this paper, we propose a progressive multi-scale consistent network for the multi-class fundus lesion segmentation from fundus images, named PMCNet. The core idea is to make full use of multi-scale feature series from adjacent layers to facilitate multi-scale feature learning and maintain feature consistency. It is implemented by adopting the proposed PFF and DAB, and they leverages high-level semantic features and low-level detailed features and enables progressive feature learning to capture rich information from both shallow and deep layers.
Experimental results on three public fundus lesion segmentation datasets demonstrate that the proposed method achieves superior performance compared with other state-of-the-art methods, which shows that consistent multi-scale feature learning in a progressive manner is helpful for the segmentation of lesions with complex contextual information in fundus images. 

In future, we will examine a semi-supervised learning strategy to make full use of fundus images without manual lesion labels, as large-scale and well-labeled data are laborious and expensive to collect, especially for pixel-level medical image analysis.
We can use a small amount of existing data with fine-grained annotation to guide the model mining potential lesion information from fundus images without pixel-level annotations, to make better use of implicit lesion information existing in a large number of fundus images.

{\small
\bibliographystyle{ieee}
\bibliography{egbib}
}

\end{document}